\documentclass[preprint]{emulateapj}

\usepackage{rotating}
\usepackage{graphicx}

\usepackage{graphicx}
\usepackage{rotating,lscape}
\usepackage{epsfig}

\def\ha{\relax \ifmmode {\rm H}\alpha\else H$\alpha$\fi}
\def\pa{\relax \ifmmode {\rm Pa}\alpha\else Pa$\alpha$\fi}
\def\arcsec{\hbox{$^{\prime\prime}$}}
\def\nii{\relax \ifmmode {\rm N\,{\sc ii}}\else N\,{\sc ii}\fi}
\def\hii{\relax \ifmmode {\rm H\,{\sc ii}}\else H\,{\sc ii}\fi}
\def\hi{\relax \ifmmode {\rm H\,{\sc i}}\else H\,{\sc i}\fi}
\def\deg{\hbox{$^{\circ}$}}

\shorttitle{The thick disk in NGC~4244}
\shortauthors{Comer\'on et al.}

\begin{document}


\title{The thick disk in the galaxy NGC~4244 from S$^4$G imaging}


\author{
S\'ebastien Comer\'on,\altaffilmark{1,2}
Johan H.~Knapen,\altaffilmark{2,3}
Kartik Sheth,\altaffilmark{4,5,6}
Michael W.~Regan,\altaffilmark{7}
Joannah L.~Hinz,\altaffilmark{8}
Armando Gil de Paz,\altaffilmark{9}
Kar\'in Men\'endez-Delmestre,\altaffilmark{10}
Juan-Carlos Mu\~noz-Mateos,\altaffilmark{9}
Mark Seibert,\altaffilmark{10}
Taehyun Kim,\altaffilmark{4,11}
E.~Athanassoula,\altaffilmark{12}
Albert Bosma,\altaffilmark{12}
Ronald J.~Buta,\altaffilmark{13}
Bruce G.~Elmegreen,\altaffilmark{14}
Luis C.~Ho,\altaffilmark{10}
 Benne W.~Holwerda,\altaffilmark{15}
Eija Laurikainen,\altaffilmark{16,17}
Heikki Salo,\altaffilmark{16}
and Eva Schinnerer\altaffilmark{18}.}

\altaffiltext{1}{Korea Astronomy and Space Science Institute, 61-1 Hwaam-dong, Yuseong-gu, Daejeon 305-348, Republic of Korea}
\altaffiltext{2}{Instituto de Astrof\'isica de Canarias, E-38200 La
Laguna, Spain}
\altaffiltext{3}{Departamento de Astrof\'isica, Universidad de La Laguna, E-38205 La Laguna, Tenerife, Spain}
\altaffiltext{4}{National Radio Astronomy Observatory / NAASC, 520 Edgemont Road, Charlottesville, VA 22903, USA}
\altaffiltext{5}{Spitzer Science Center}
\altaffiltext{6}{California Institute of Technology, 1200 East California Boulevard, Pasadena, CA 91125, USA}
\altaffiltext{7}{Space Telescope Science Institute, 3700 San Martin Drive, Baltimore, MD 21218, USA}
\altaffiltext{8}{Steward Observatory, University of Arizona, 933 North Cherry Avenue, Tucson, AZ 85721, USA}
\altaffiltext{9}{Departamento de Astrof\'isica, Universidad Complutense de Madrid, 28040, Madrid, Spain}
\altaffiltext{10}{The Observatories of the Carnegie Institution of Washington, 813 Santa Barbara Street, Pasadena, CA 91101, USA}
\altaffiltext{11}{Astronomy Program, Department of Physics and Astronomy, Seoul National University, Seoul 151-742, Republic of Korea}
\altaffiltext{12}{Laboratoire d'Astrophysique de Marseille (LAM), UMR6110, CNRS/Universit\'e de Provence/CNRS, Technop\^ole de Marseille \'Etoile, 38 rue Fr\'ed\'eric Joliot Curie, 13388 Marseille C\'edex 20, France}
\altaffiltext{13}{Department of Physics and Astronomy, University of Alabama, Box 870324, Tuscaloosa, AL 35487, USA}
\altaffiltext{14}{IBM T.~J.~Watson Research Center, 1101 Kitchawan Road, Yorktown Heights, NY 10598, USA}
\altaffiltext{15} {Astrophysics, Cosmology and Gravity Centre (ACGC), Astronomy Department, University of Cape Town, Private Bag X3, 7700 Rondebosch, Republic of South Africa}
\altaffiltext{16}{Astronomy Division, Department of Physical Sciences, P.~O.~Box 3000, FIN-90014 University of Oulu, Finland}
\altaffiltext{17}{Finnish Centre of Astronomy with ESO (FINCA), University of Turku, V\"ais\"al\"antie 20, FI-21500, Piikki\"o, Finland}
\altaffiltext{18}{Max-Planck-Institut f\"ur Astronomie, K\"onigstuhl 17, D-69117 Heidelberg, Germany}


\begin{abstract}

 If thick disks are ubiquitous and a natural product of disk galaxy formation and/or evolution processes, all undisturbed galaxies which have evolved during a significant fraction of a Hubble time should have a thick disk. The late-type spiral galaxy NGC~4244 has been reported as the only nearby edge-on galaxy without a confirmed thick disk. Using data from the {\it Spitzer Survey of Stellar Structure in Galaxies} (S$^{4}$G) we have identified signs of two disk components in this galaxy. The asymmetries between the light profiles on both sides of the mid-plane of NGC~4244 can be explained by a combination of the galaxy not being perfectly edge-on and a certain degree of opacity of the thin disk. We argue that the subtlety of the thick disk is a consequence of either a limited secular evolution in NGC~4244, a small fraction of stellar material in the fragments which built the galaxy, or a high amount of gaseous accretion after the formation of the galaxy.

\end{abstract}




\section{Introduction}

Thick disks are detected in edge-on and very inclined galaxies as an excess of flux at some distance from the mid-plane, typically at a few thin disk scale-heights. They usually have exponential profiles with a larger scale-height than the `canonical' or thin disks. Thick disks were first detected by Tsikoudi (1979) and were defined by Burstein (1979). A thick disk has subsequently been found in the Milky Way (Gilmore \& Reid 1983) and in basically all kinds of disk galaxies (see the introduction of Yoachim \& Dalcanton 2008 and references therein).

Several theories of thick disk formation have been proposed. It has been suggested that thick disks are formed through mechanisms of internal evolution, namely disk heating by gravitational instabilities (Bournaud et al.~2007; 2009) and/or stellar scattering by encounters with features such as dark matter subhaloes, spiral arms or giant molecular clouds (Villumsen 1985; H\"anninen \& Flynn 2002; Benson et al.~2004; Hayashi \& Chiba 2006; Haywood 2008; Kazantzidis et al.~2008; Sch\"onrich \& Binney 2009a; Sch\"onrich \& Binney 2009b ; Bournaud et al.~2009). In this picture the thick disk formation may take less than one gigayear in the case of fast dynamical heating by a very clumpy initial disk (Bournaud et al.~2009) or may be much slower (secular evolution; Villumsen 1985; H\"anninen \& Flynn 2002; Benson et al.~2004; Kazantzidis et al.~2008). The specific type of secular evolution which causes a slow heating of the stellar orbits has been termed `blurring' by Sch\"onrich \& Binney (2009a; 2009b). 

Another possibility is that the thick disk is formed through accretion of stellar material from merging satellites at the time of galaxy formation (Abadi et al.~2003; Yoachim \& Dalcanton 2006). A third possibility is that the thick disk is a consequence of {\it in situ} star formation (Brook et al.~2004, Elmegreen \& Elmegreen 2006) or of star formation with a high initial velocity dispersion in very massive star clusters (Kroupa 2002).

Whatever the thick disk formation mechanism is, it has to act over most, if not all, disk galaxies as thick disks have been found to be ubiquitous (Dalcanton \& Bernstein 2002; Yoachim \& Dalcanton 2006). Therefore, if any disk galaxy was found not to possess a thick disk, it would be likely to have suffered a peculiar origin and/or evolution. A search in the literature yields a candidate, NGC~4244, which has been claimed either to be a doubtful case (Yoachim \& Dalcanton 2006), or not to have a thick disk at all (Fry et al.~1999). NGC~4244 is a highly inclined apparently undisturbed Scd galaxy located 4.4\,Mpc away (Seth et al.~2005a). At that distance, one arcsecond is equivalent to roughly 21\,pc in linear scale. NGC~4244 is nearly bulgeless, except for a compact and bright nuclear star cluster (see Seth et al.~2008 for a detailed study on the nuclear cluster; the bulge-to-disk ratio is 0.03 according to Fry et al.~1999). NGC~4244 exhibits systematic displacements of the mid-plane of a few tens of parsecs with a wave-length of a couple of kiloparsecs (this phenomenon has been termed `corrugation' and in the case of NGC~4244 it has been studied by Florido et al.~1991). Kodaira \& Yamashita (1996) found NGC~4244 to have less warm dust and star formation than galaxies of the same Hubble type and \hi\ content, thus labeling it as an `anemic' galaxy. Finally, de Jong et al.~(2007) found a sharp disk cut-off which they suggest to have a dynamical origin. NGC~4244 could be weakly interacting with NGC~4214, which has a similar recession velocity and is found at a projected distance of 120\,kpc from NGC~4244.

Several studies have attempted to identify the structural components of NGC~4244. Bergstrom et al.~(1992) found no evidence for a massive stellar halo. Fry et al.~(1999) detected only a thin disk. Hoopes et al.~(1999) found that, unlike some other edge-on galaxies, NGC~4244 has its ionized gas concentrated in the thin disk. Strickland et al.~(2004) found no evidence for a radio, UV, or X-ray emitting halo. Seth et al.~(2005b) found that the scale-heights of the older stellar components are larger than those of the younger components, which could be an indicator of the existence of a thick disk; however, the difference in scale-heights between the main sequence population and the red giant branch (RGB) population is smaller than in all the other galaxies studied in their paper. They also detected what could be interpreted as a halo. Tikhonov \& Galazutdinova (2005) claimed to have discovered a tenuous extended halo with a scale-length of several kiloparsecs in NGC~4244. Seth et al.~(2007) found a faint halo component 2.5\,kpc above the mid plane at the position of the minor axis.

To sum up, previous studies on NGC~4244 show little to no emission at any wavelength outside its well-known thin disk. As nearly all very inclined disk galaxies have been shown to host a thick disk, the absence of it in NGC~4244, if confirmed, would be an indication of a peculiar evolution or formation mechanism of this galaxy.

In this Paper we present results of a study of NGC~4244 based on a {\it Spitzer Survey of Stellar Structure in Galaxies} (S$^{4}$G) $3.6\mu{\rm m}$ image (Sheth et al.~2010; Sect.~2), confirming that this galaxy does in fact have signs of two disk components. For this purpose we have fitted the galaxy using {\sc Galfit} (version~3.0; Peng et al.~2002; Peng et al.~2010; Sect.~3) and produced luminosity profiles of NGC~4244 at different galactocentric radii (Sect.~4). We describe the luminosity profiles and compare them with a simple inclined two-disk model with a diffuse dust contribution in the thin disk in Sect.~5. We present a summary of the NGC~4244 main structural components in Sect.~6, and discuss our findings in Sect.~7.

\section{Processing of the S$^{4}$G $3.6\mu{\rm m}$ image of NGC~4244}

\begin{figure*}[!t]
\begin{center}
\begin{tabular}{c}
\includegraphics[width=0.45\textwidth]{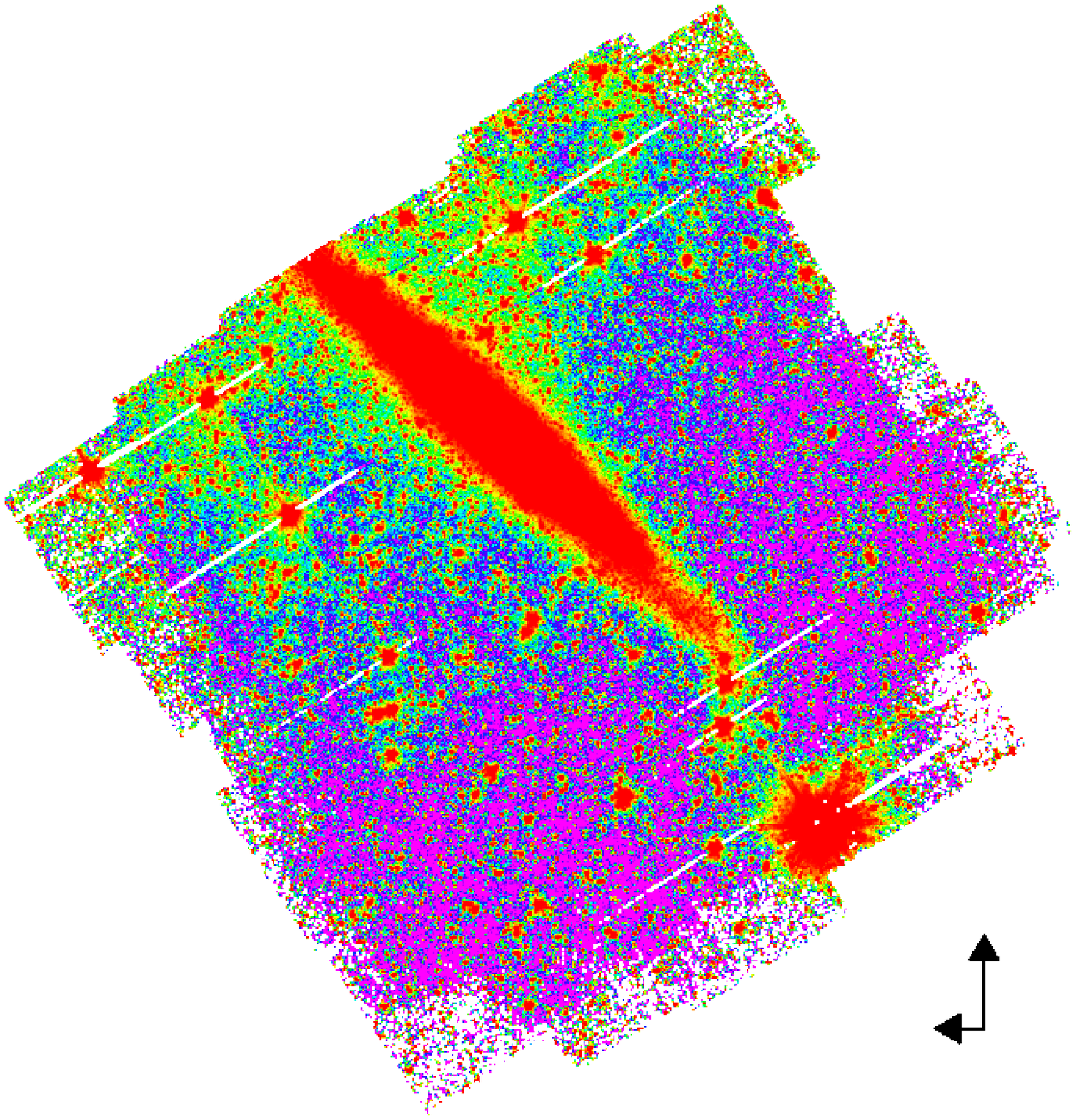}
\includegraphics[width=0.45\textwidth]{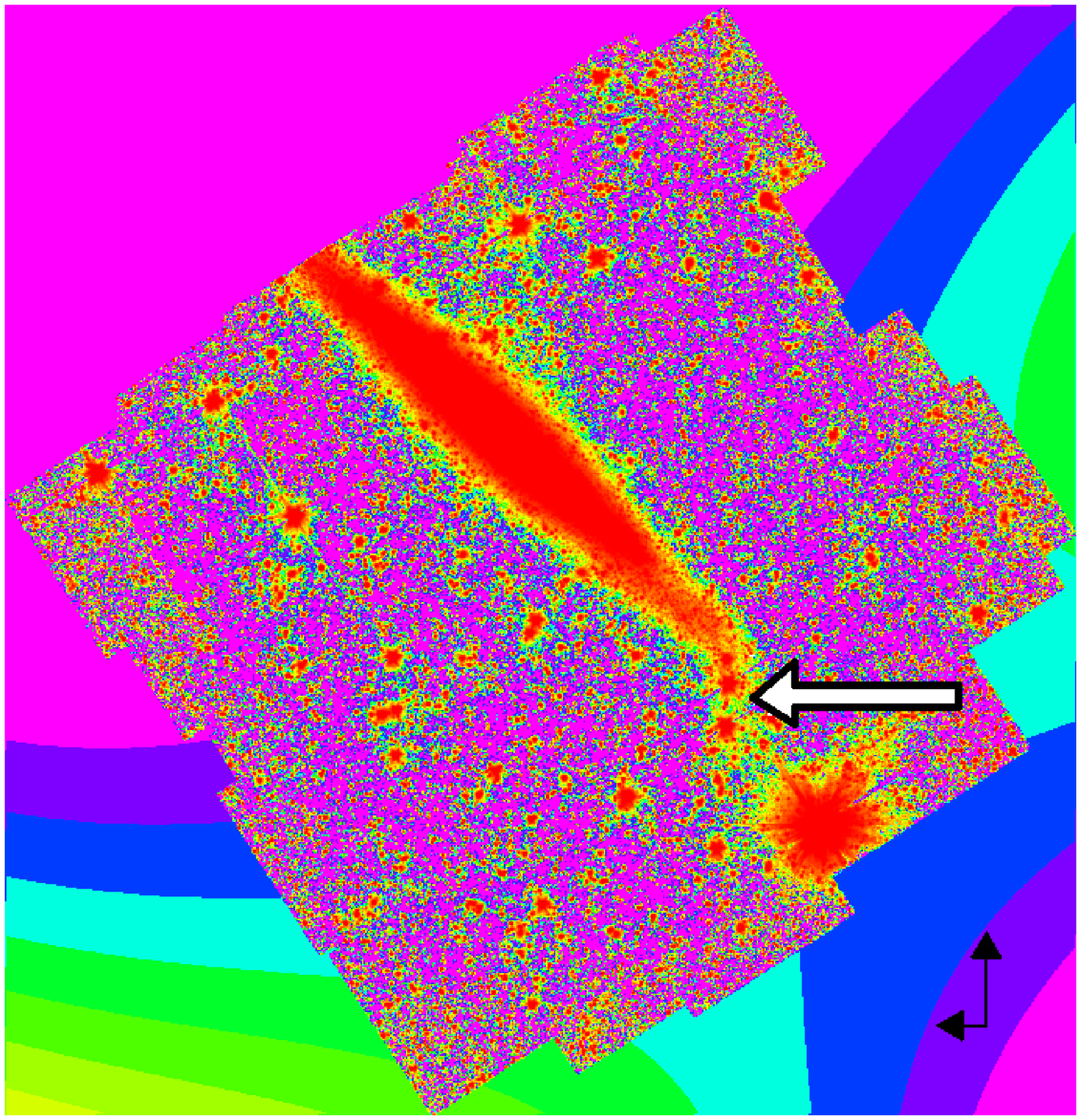}\\
\includegraphics[width=0.91\textwidth]{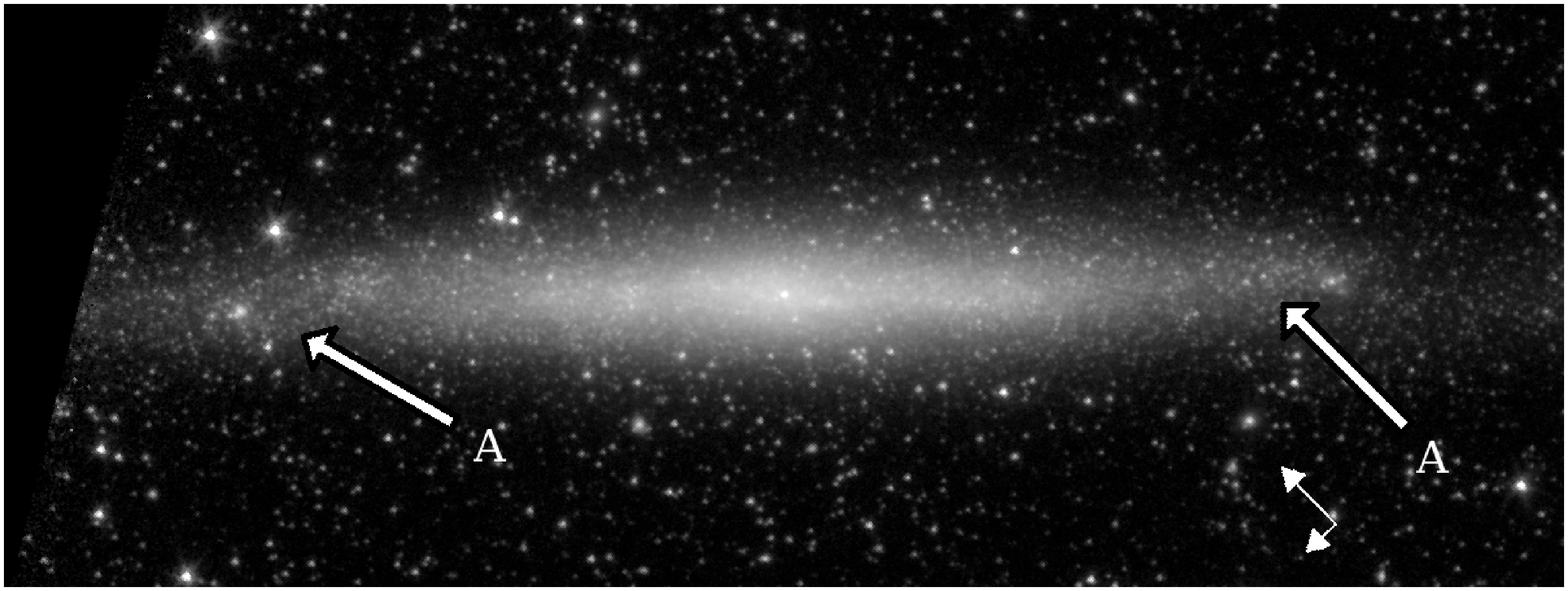}\\
\end{tabular}
\caption{\label{images4244} Top panels: the $3.6\mu{\rm m}$ S$^{4}$G image of NGC~4244 before (left) and after (right) removing the gradient in the background. We have used the same histogram equalization display for both images. These images have been smoothed using a Gaussian function with $\sigma=1.25$\,pixels. North is up and East is left. Lower panel: $3.6\mu{\rm m}$ S$^{4}$G image of NGC~4244 rotated 43.6\deg counter-clockwise in such a way that the galaxy plane is horizontal. In this frame North is roughly located at the top-left and East at the bottom-left. The image is 15\arcmin\ or 19.2\,kpc in length. The distance between the nuclear cluster and the `A' arrows is around 5\arcmin. The display for this panel is logarithmic. The arrow in the top-right panel indicates a light streak caused by the combination of the light of three foreground stars. The `A' arrows in the bottom panel indicate the location of the warp-like feature.}
\end{center}
\end{figure*}

The S$^{4}$G, described by Sheth et al.~(2010), is a survey using the 3.6$\,\mu$m and 4.5$\,\mu$m filters of the {\it Spitzer Space Telescope} Infrared Array Camera (IRAC, Fazio et al.~2004) with the aim of studying the stellar mass distribution in a sample of 2331 local galaxies.

The $3.6\mu{\rm m}$ image of NGC~4244 we have used for our study has been processed by the S$^{4}$G pipelines, resulting in a mosaicked flat-fielded image and in a mask of foreground stars, background galaxies and bad pixels (Sheth et al.~2010). However, as we aim to reach the faintest possible surface brightness level, we have manually refined the mask and closely examined the background. The background of the image is at a level of $0.0330\,{\rm mJy\,sr^{-1}}$ (corresponding to $\mu_{3.6\mu{\rm m}}(AB)=24.2\,{\rm mag\,arcsec^{-2}}$). The $\sigma$ of the background is of the order of $0.002\,{\rm mJy\,sr^{-1}}$, which is comparable to the offsets applied to the individual frames in order to obtain a mosaic with a uniform background ($0.0015\,{\rm mJy\,str^{-1}}$). We find a systematic background gradient which is shown in the top-left panel in Fig.~\ref{images4244}. This gradient has an amplitude which is approximately two times larger than that of the $\sigma$ of the background. The gradient is unlikely to be caused by zodiacal light or by the concentration of bright stars at the northern part of the frame because it is not consistent with that of the S$^{4}$G $4.5\mu{\rm m}$ image which was obtained for the same galaxy. We have modeled the background using a third order two-dimensional polynomial fit to all unmasked pixels at a distance of 250 pixels (187.5\arcsec) or more from the galaxy mid-plane. In this procedure there is no risk of contamination by any component of the galaxy, because we are going farther than any trace of extended component found by Tikhonov \& Galazutdinova (2005) using the RGB star-counting method and because the surface brightness of the halo detected by Seth et al.~(2007) drops below the S$^{4}$G limiting magnitude at that distance from the mid-plane. We then subtracted the modeled background from the original image (the result is shown in the top-right panel in Fig.~\ref{images4244}).

\begin{figure*}[!t]
\begin{center}
\begin{tabular}{c}
\includegraphics[width=0.91\textwidth]{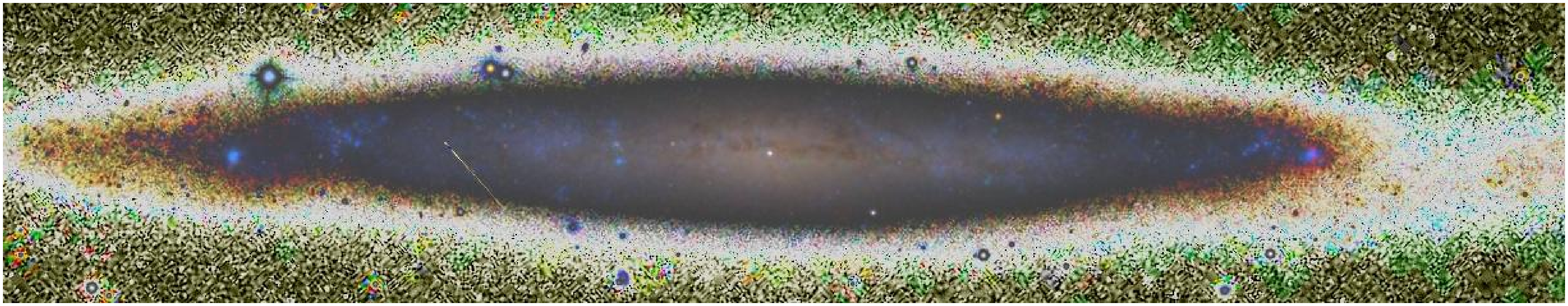}\\
\includegraphics[width=0.91\textwidth]{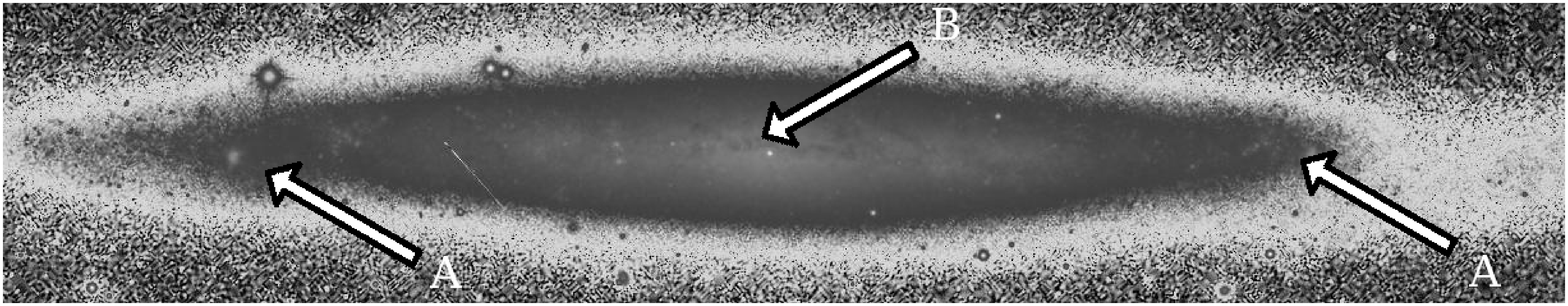}
\end{tabular}
\caption{\label{dr7} SDSS DR7 $ugriz$ image of NGC~4244 oriented identically and at the same scale as in the bottom panel of Fig.~\ref{images4244}. The top panel contains a color version of the image and the bottom panel has a black and white version of it. In the color image, the blue knots denote star formation and the reddish colors dusty areas. In the bottom panel, the `A' arrows indicate the location of the warp-like feature which is described in Sect.~2. The `B' arrow indicates a dust lane that runs parallel to the mid-plane of the galaxy and is located slightly above it.}
\end{center}
\end{figure*}

We rotated the image $43.6\deg$ counter-clockwise in order to place the mid-plane of the galaxy horizontally (lower panel in Fig.~\ref{images4244}). We estimate that the uncertainty in the rotation angle is about one degree. The rotation angle has been calculated by finding the mid-plane for all galactocentric distances $r<350\,{\rm arcsec}$ ($r<7.5$\,kpc). This rotation angle is within the uncertainties consistent with the value of $42\deg$ used by Olling (1996) and Fry et al.~(1999). Once the galaxy has been rotated, we can define as the {\it top} part of the galaxy the area extending above its mid-plane and as the {\it bottom} part of the galaxy the region below it. This definition is arbitrary, but it will be useful because the galaxy is not symmetric with respect to its mid-plane.

NGC~4244 has a warp-like feature starting from galactocentric radii of 4 to 5\arcmin\ in both optical (indicated with an arrow in Fig.~\ref{dr7}; SDSS DR7 image; Abazajian et al.~2009) and near-infrared bands (indicated with arrows in the bottom panel in Fig.~\ref{images4244}) in agreement with the warp properties measured by Saha et al.~(2009). The color image of the galaxy (top panel in Fig.~\ref{dr7}) shows that this warp-like feature is delineated by a succession of blue knots [$\left(g_{\rm knot}-r_{\rm knot}\right)-\left(g_{\rm surroundings}-r_{\rm surroundings}\right)\sim-0.4$], which are areas of ongoing star formation as confirmed by the \ha\ image of Kennicutt et al.~(2008). This indicates that the warp-like feature is likely to be a product of the presence of spiral arm fragments embedded in a very inclined galaxy disk (this possibility has been discussed by van der Kruit 1979). In the bottom panel of Fig.~\ref{dr7} the warp-like feature is delineated by the deviations of the tips of the gray areas (indicated with `A' arrows) from the horizontal orientation. This warp-like feature is part of the corrugation pattern described by Florido et al.~(1991), as outside the `warp', at a lower surface brightness, the disk mid-plane bends in the other sense (seen in Fig.~\ref{dr7}). In the top images of Fig.~\ref{images4244} (in which the color-scale distribution is based on a histogram equalization algorithm), the warp-like feature we have described may be confused with a light streak which goes downward from the south-eastern edge of NGC~4244 (opposite to the warp-like feature and indicated by an arrow). This streak is caused by three relatively bright aligned stars whose PSFs overlap in the histogram equalized display.

The warp-like distortion we find coincides with that seen in the $R$-band by Fry et al.~(1999). These authors state that the optical `warp' they see is coincident with the \hi\ warp discussed by Olling (1996), but this conclusion is due to them having used an image of NGC 4244 flipped along the mid-plane. The warp found in \hi\ by Olling (1996) starts at a galactocentric radius of $r\sim10$\arcmin, which would place it at larger galactocentric radii than shown in the image in the bottom panel of Fig.~\ref{images4244}.

An inspection of the SDSS DR7 image of NGC~4244 (Fig.~\ref{dr7}; indicated by the arrow labeled with letter `B') shows a dust lane parallel to the mid-plane and slightly above it. This indicates that what we have defined to be the bottom part of the galaxy is facing us and thus we can describe it as the near side of the galaxy. This was already pointed out by Olling (1996), who found that the inclination of the galaxy is $i=84.5\deg$. The dust lane is undetectable in the S$^{4}$G images, even in a $3.6\,\mu{\rm m}-4.5\,\mu{\rm m}$ color-index map.

\section{GALFIT analysis of NGC~4244}

\begin{figure*}[!t]
\begin{center}
\begin{tabular}{c c c}
&Inclined single-disk model & Edge-on single-disk model\\
\begin{sideways}Original image\end{sideways}&
\includegraphics[width=0.45\textwidth]{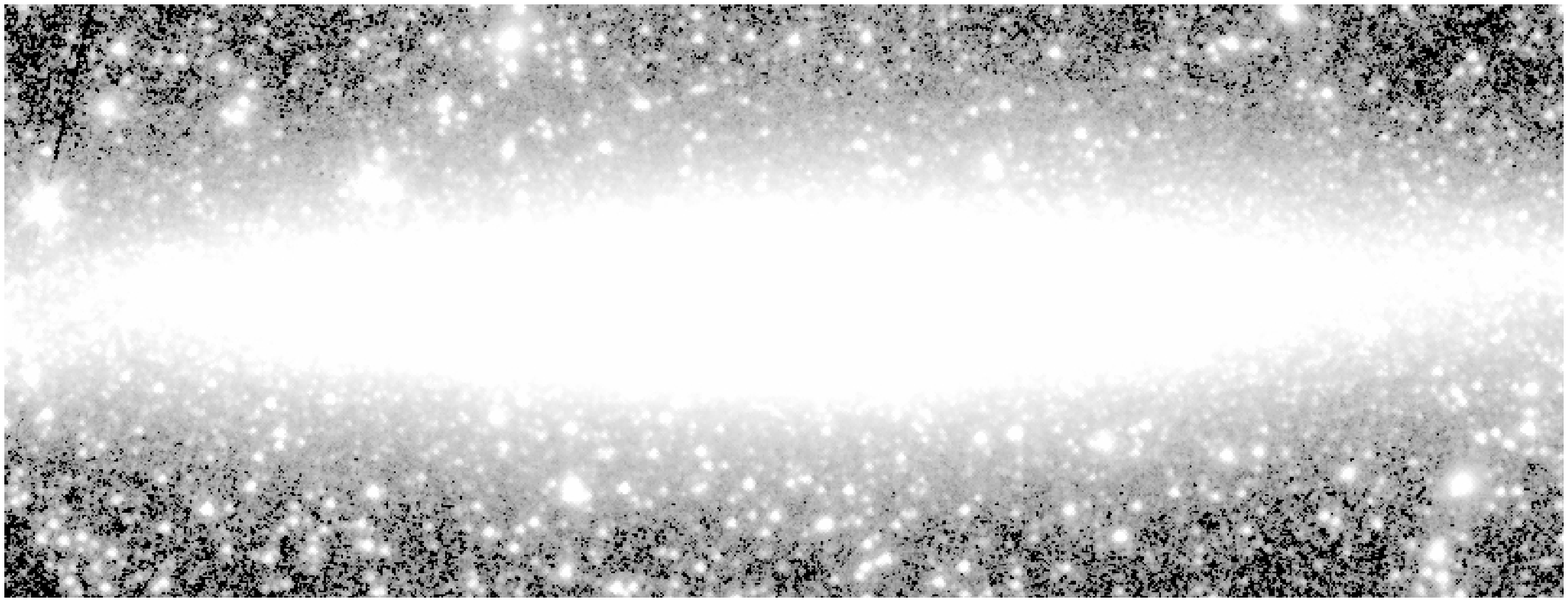}&
\includegraphics[width=0.45\textwidth]{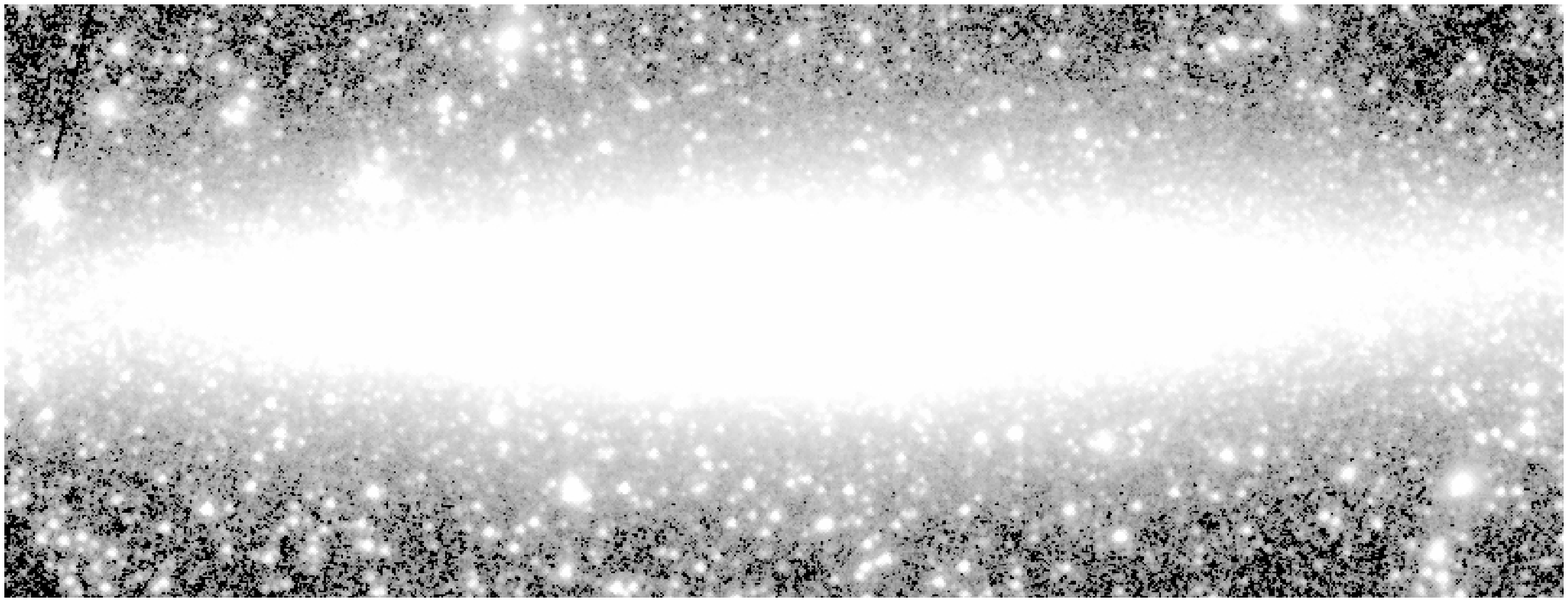}\\
\begin{sideways}Model\end{sideways}&
\includegraphics[width=0.45\textwidth]{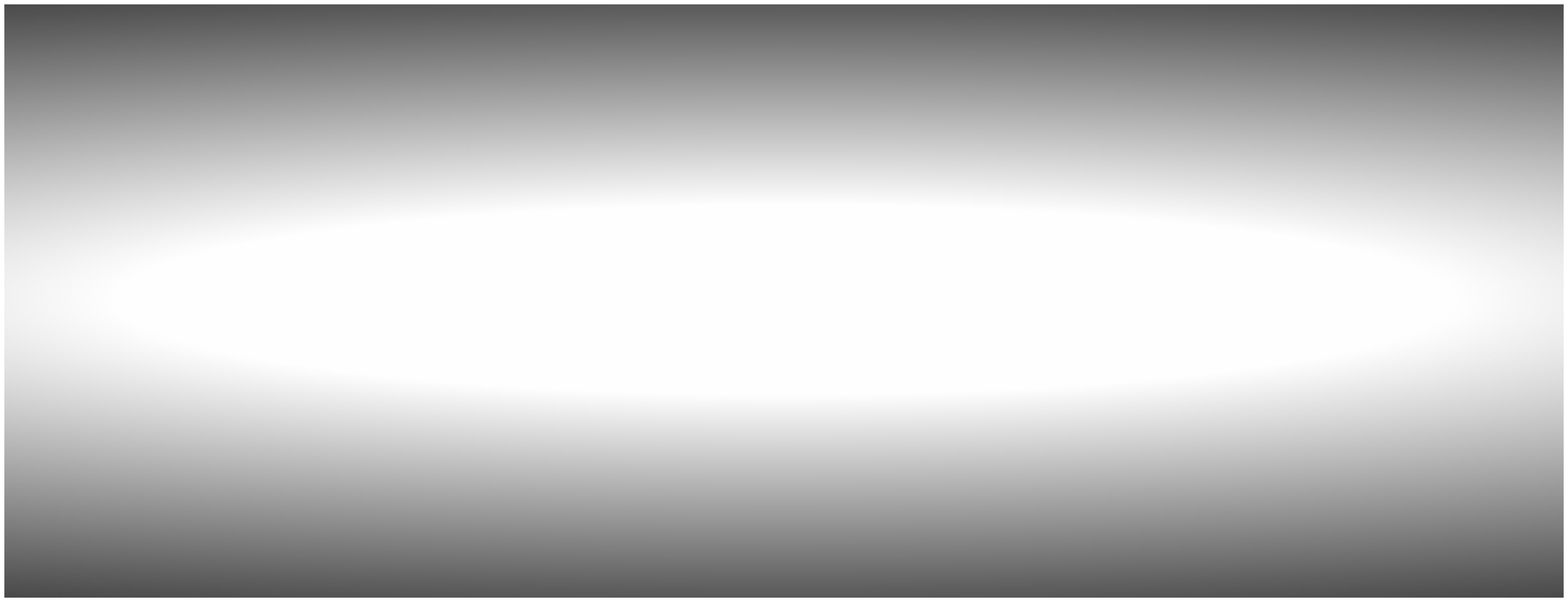}&
\includegraphics[width=0.45\textwidth]{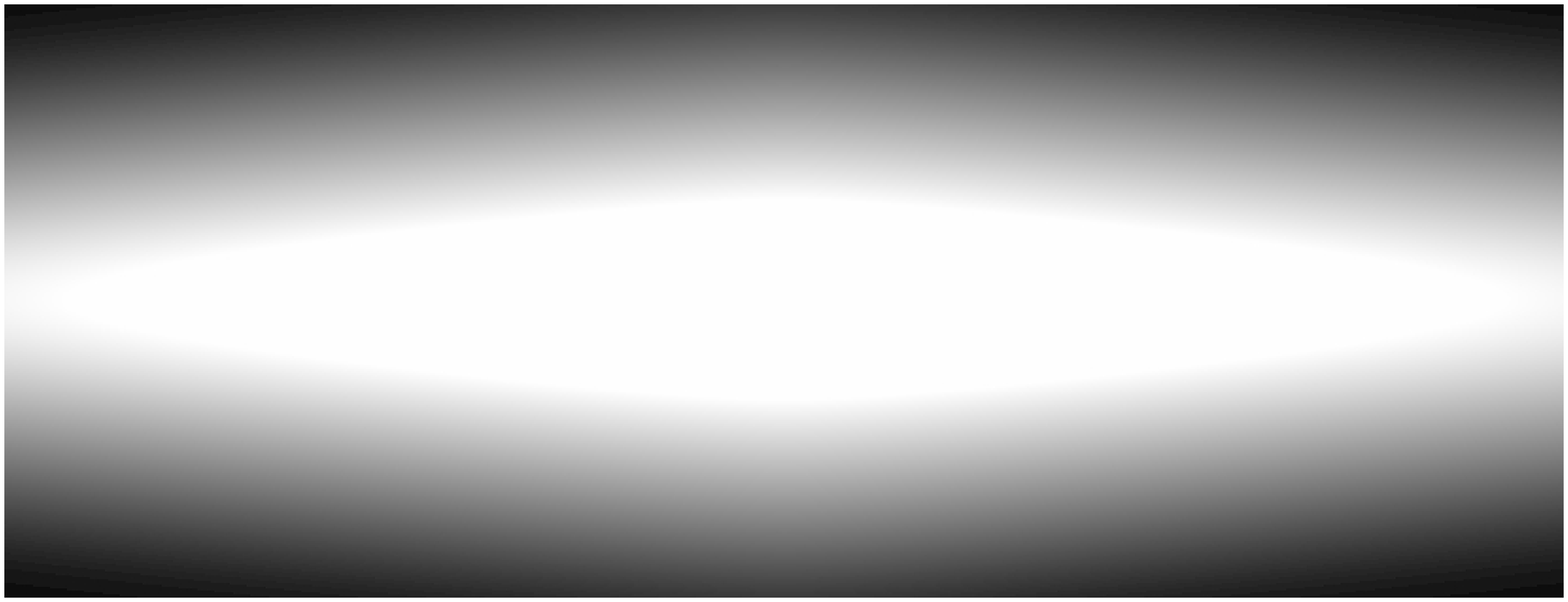}\\
\begin{sideways}Residual\end{sideways}&
\includegraphics[width=0.45\textwidth]{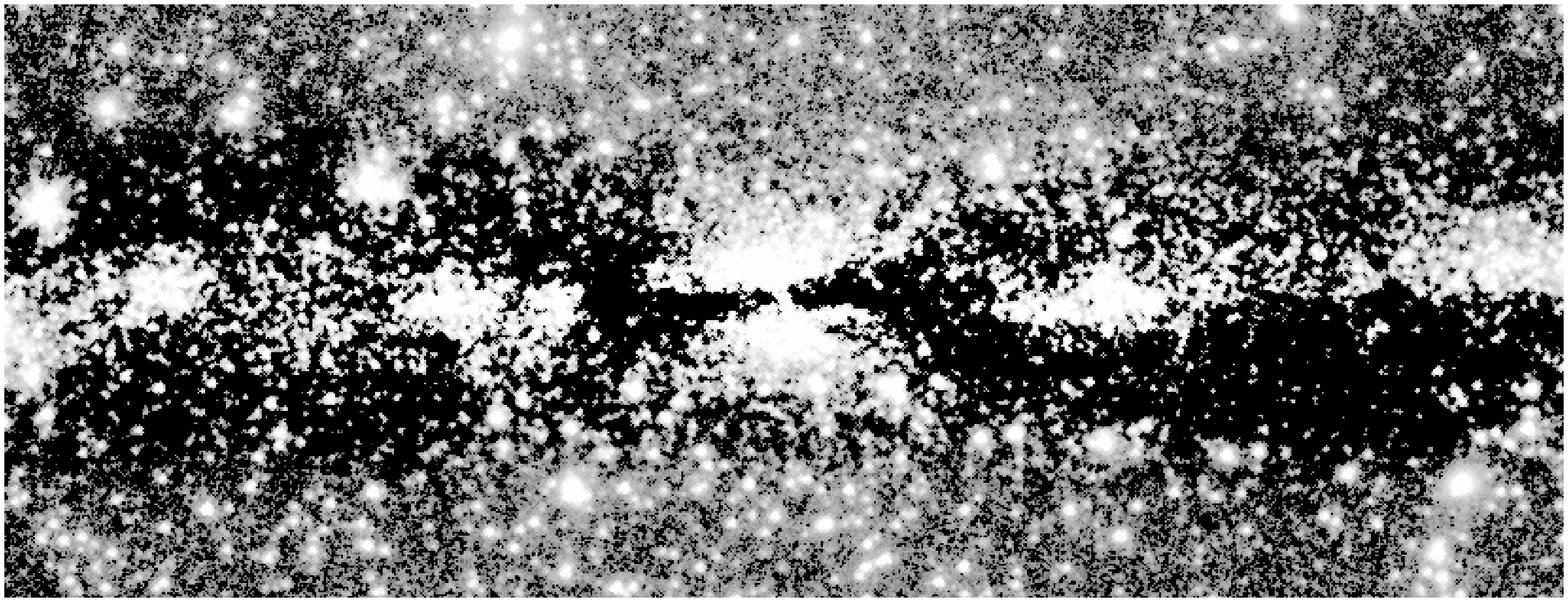}&
\includegraphics[width=0.45\textwidth]{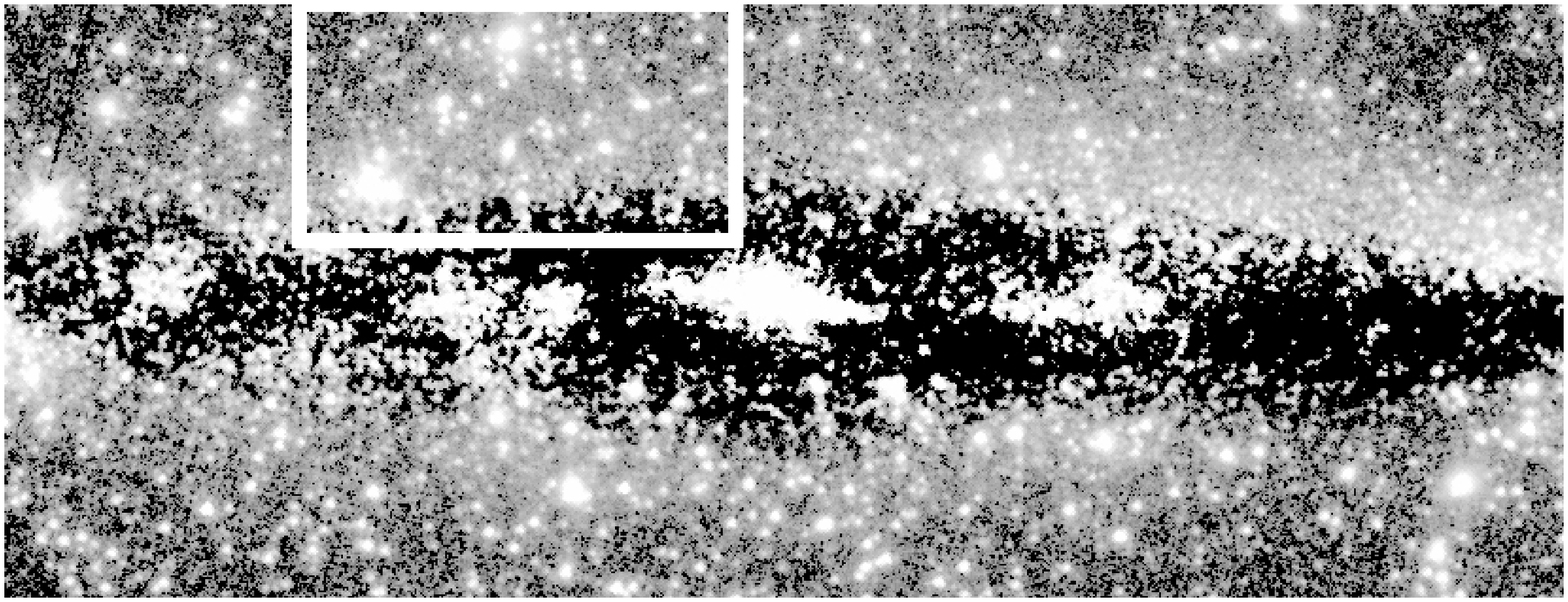}\\
\end{tabular}
\caption{\label{modelsngc4244} The inclined single-disk model is presented on the left and the edge-on single-disk model is presented on the right. From top to bottom, the images correspond to the original S$^{4}$G background-subtracted frame, to the {\sc Galfit} model, and to the residual image. All the images are in a logarithmic display. The white rectangle indicates an arc-like feature that we interpret as possibly of tidal origin.}
\end{center}
\end{figure*}

\begin{figure}[!t]
\begin{center}
\begin{tabular}{c}
\includegraphics[width=0.45\textwidth]{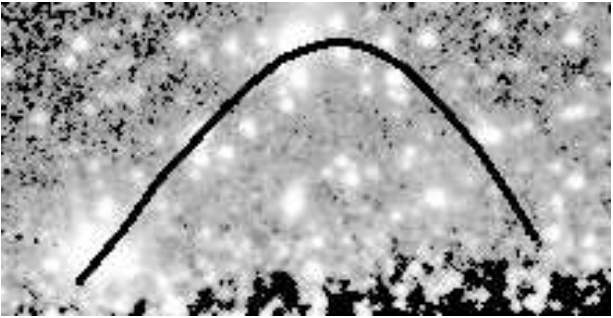}\\
\end{tabular}
\caption{\label{arc} Close-up of the box in the lower-right panel in Fig.~\ref{modelsngc4244}. The proposed path of the arc-like feature is indicated with a black line.}
\end{center}
\end{figure}

In order to know whether a single-disk description is good enough to account for the structural components of NGC~4244 we have run {\sc Galfit} (version 3.0; Peng et al.~2002; Peng et al.~2010) on the S$^{4}$G NGC~4244 image using the {\sc Galfidl} procedures and display routines (see http://sun3.oulu.fi/~hsalo/galfidl.html). {\sc Galfit} is a code designed for fitting one or more two-dimensional components to an image.

As an input we used the background-subtracted S$^4$G image, the refined mask, and a `sigma' image, which accounts for the statistical uncertainties associated to each pixel. The `sigma' image is made using {\sc Galfidl}, which takes into account the pixel weights varying in different parts of the image mosaic.

\subsection{Inclined disk model}

In this model {\sc Galfit} assumes a radially decaying exponential profile. In order to run it we fixed the center of the galaxy at the position of the nuclear cluster and the disk PA as 46.4\deg\ which corresponds to 90\deg\ in the rotated images we present in Fig.~\ref{modelsngc4244}. We left as free parameters the disk luminosity, the disk axis ratio, and the disk scale-length. As the galaxy appears to contain no bulge, no further components are {\it a priori} needed. The major caveat of modeling a very inclined galaxy this way is that the inclined disk {\sc Galfit} model is an infinitely thin circular disk model, which will cause all the vertical extension of the galaxy to be interpreted as caused by the disk inclination. The resulting disk has an axis ratio $d/D=0.14$, which gives a disk inclination of $i=82\deg$. Considering that in our model the disk is infinitely thin, which will automatically cause the disk to be less inclined than it really is, this value is compatible with the inclination of $i=84.5\deg$ given by Olling (1996). The inclined disk model and the residual image (the original background-subtracted image from which the model has been subtracted) can be seen in the left column of Fig.~\ref{modelsngc4244}.

\subsection{Edge-on disk model}

When applying the edge-on disk {\sc Galfit} model we are assuming that the inclination angle of the galaxy is $i=90\deg$. The function which {\sc Galfit} uses is 
\begin{equation}
\Sigma(r,z)=\Sigma_0\frac{|r|}{h_{\rm r}}K_1\left(\frac{|r|}{h_{\rm r}}\right){\rm sech^2}\left(\frac{z}{h_{\rm z}}\right)
\end{equation}
\noindent where $\Sigma_0$ is the central surface brightness, $K_1$ is a Bessel function, $r$ is the galactocentric distance, $h_{\rm r}$ is the major axis disk scale-length, $z$ is the distance above the plane, and $h_{\rm z}$ is the scale-length perpendicular to the plane, as developed by van der Kruit \& Searle (1981) for the light distribution in edge-on disks. As for the other model, we fixed the center of the galaxy at the position of the nuclear cluster and the disk PA as 90\deg. We left as free parameters the disk luminosity, the disk scale-length, and the disk scale-height. The major caveat of modeling a not fully edge-on galaxy this way is that we are ignoring the fact that part of the apparent scale-height is an effect of the galaxy inclination. The edge-on disk model and the residual image can be seen in the right column of Fig.~\ref{modelsngc4244}.

\subsection{Comparison of the two NGC~4244 {\sc Galfit} models}

The main difference between the inclined and the edge-on model (see Fig.~\ref{modelsngc4244}) is that in the first case the galaxy model isophotes are elliptical and that in the second case they are diamond-shaped. The reason for this is that, in the first case, we are seeing the projection of an infinitely thin disk, which gives as a result a very eccentric ellipse. In the second case we are seeing a diamond shape because the {\sc Galfit} light distribution for an edge-on disk has a cusp at the galactocentric distance $r=0$. The real galaxy image is more peaked than the inclined disk model for galactocentric distances larger than $r=4\arcmin$ (which causes the butterfly-shaped dark areas in the residual image) but has rounder isophotes than the edge-on model near $r=0$. The residual image shows that, again, the inclined disk model is inaccurate at high galactocentric radii where it predicts a light distribution parallel to the $z$-axis much wider than observed.

In both models, the residual image shows bright patches close to the mid-plane. In the residuals of the edge-on model, these patches are reminiscent of a small bar or/and a spiral arm system seen nearly edge-on. The residual image of the edge-on model also shows an envelope of residual light which may be related to a thick disk. The extension of this envelope roughly corresponds to that of the galaxy's white-color envelope in the color panel in Fig.~\ref{dr7}. However, part of this residual emission, especially that at the top-right side of the image, has more to do with the warp-like feature described in Sect.~2 than with the presence of a thick disk, as indicated in Fig.~\ref{dr7} by the behavior of blue knots of star formation at high galactocentric radii.

An arc-like feature is visible in the residual images of both models at the top-left side of the frame and at a very low surface brightness (Figs.~\ref{modelsngc4244} and~\ref{arc}). It is more visible in the edge-on model residual image (indicated with a white rectangle) and we speculate that it could be related to some kind of tidal feature, as its shape is somewhat similar to that observed in other very inclined galaxies (such as in NGC~5907 studied by Shang et al.~1998 and Mart\'inez-Delgado et al.~2008 and in NGC~4013 by Mart\'inez-Delgado et al.~2009; these cases are admittedly much more spectacular than that described here). This feature is knotty, which can be explained by the presence of globular clusters or star-forming regions in a tidally disrupted galaxy.

Neither of the models we ran reflect accurately the behavior of the stellar components in NGC~4244. However, the simple modeling we have done is enough to show that a simple single-component description is not adequate to describe NGC~4244's structural properties.

\begin{figure*}[!ht]
\begin{center}
\begin{tabular}{c|c}
Near side (bottom side) & Far side (top side) \\
&\\
\includegraphics[width=0.45\textwidth]{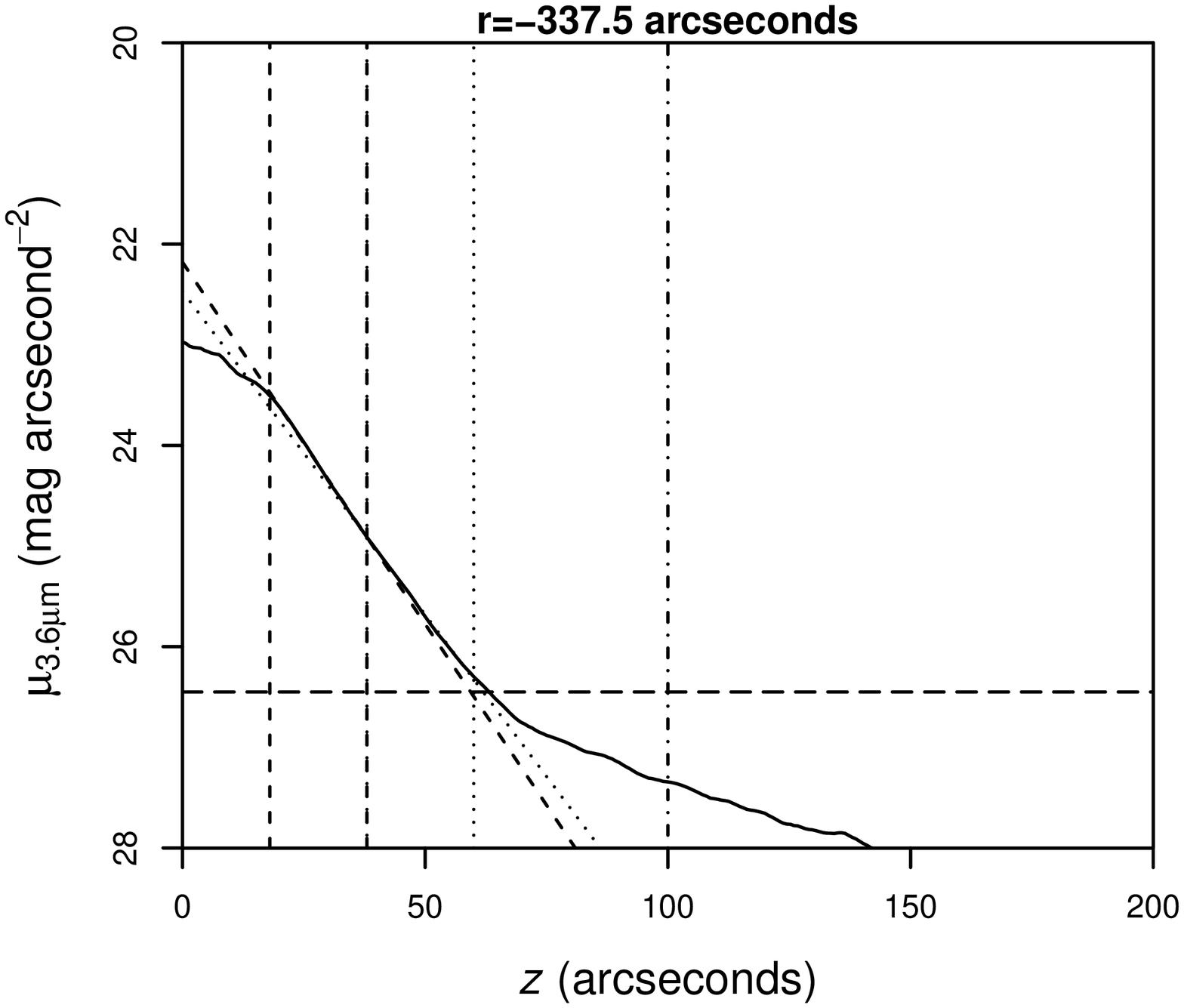}&
\includegraphics[width=0.45\textwidth]{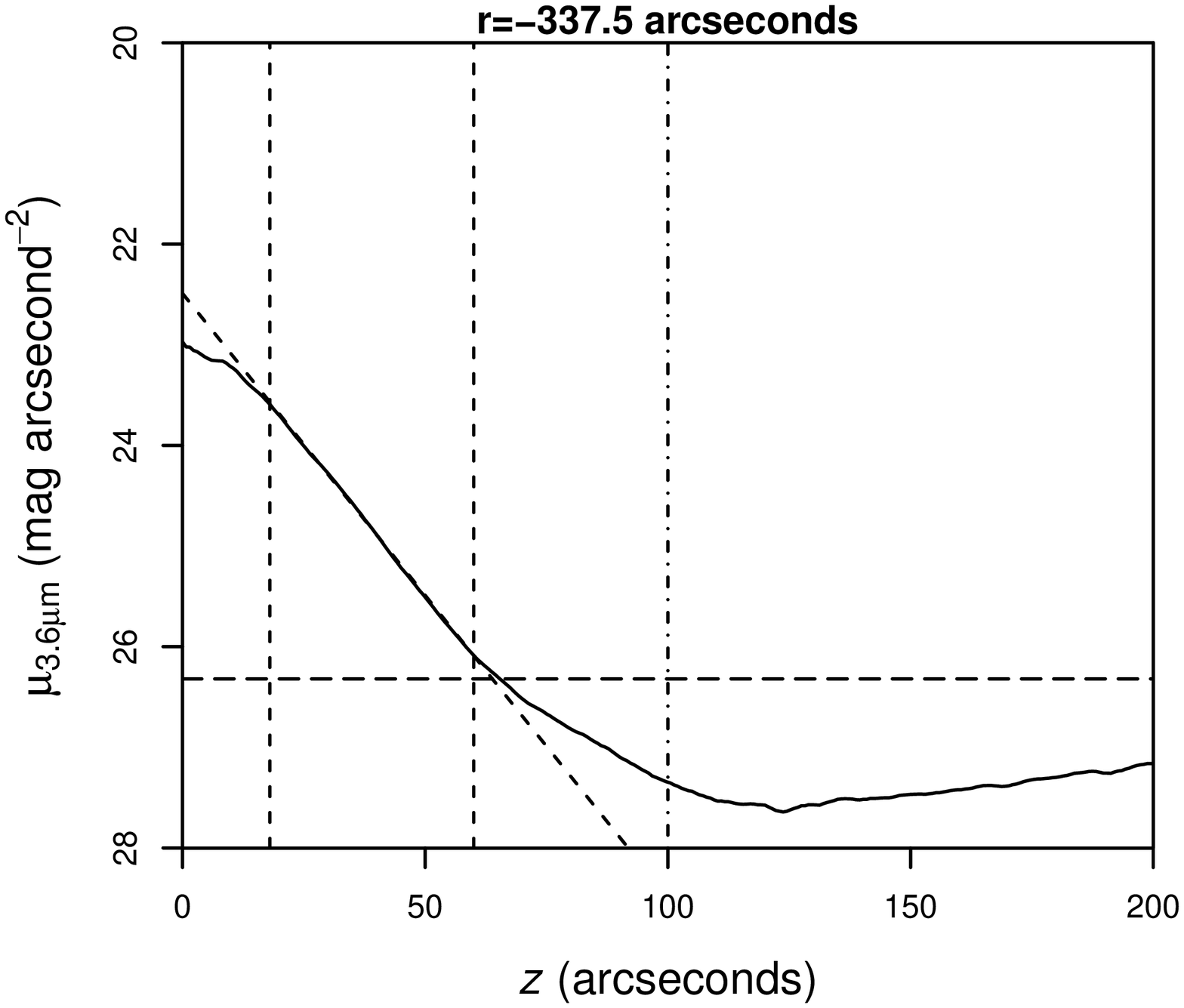}\\
\includegraphics[width=0.45\textwidth]{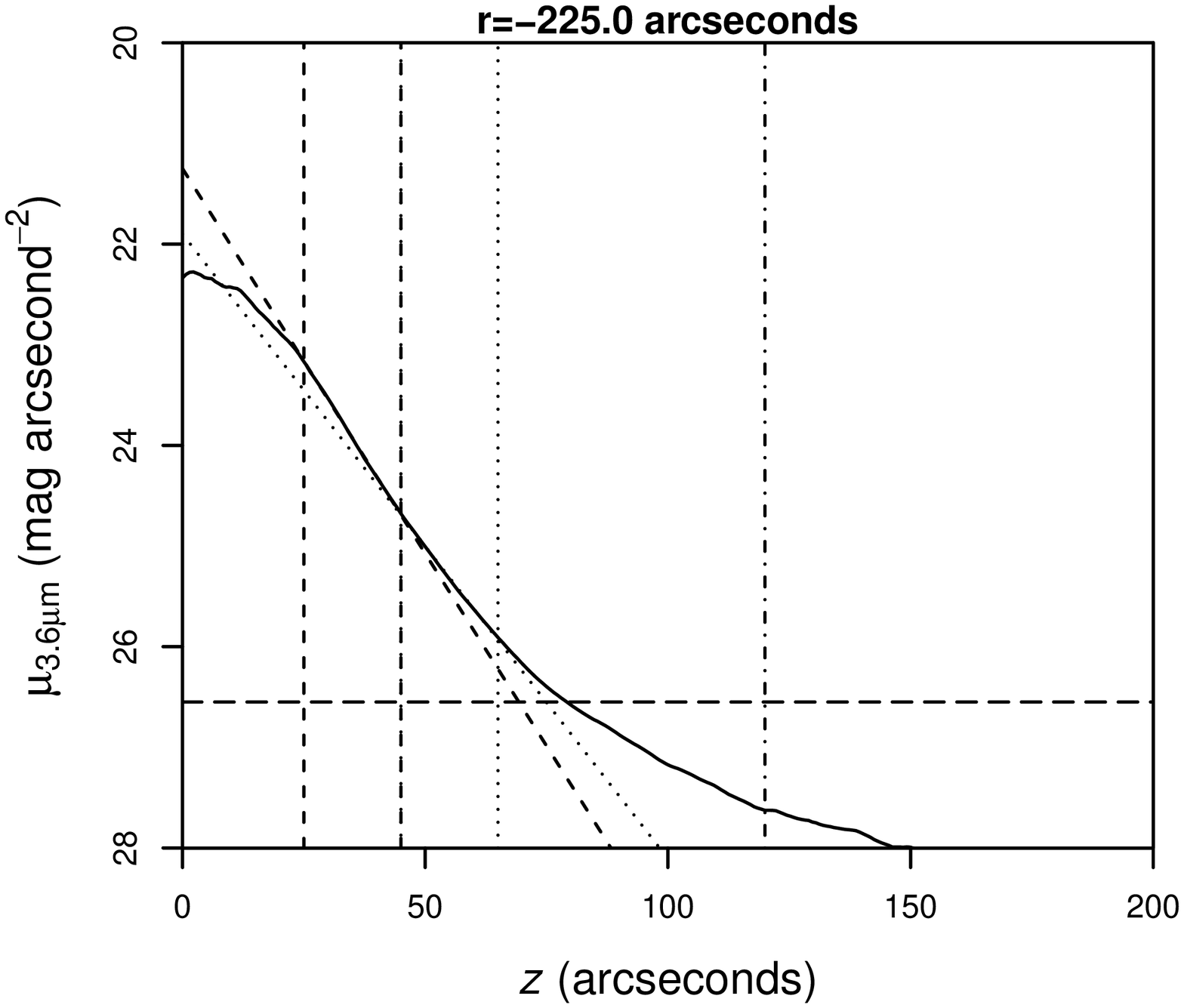}&
\includegraphics[width=0.45\textwidth]{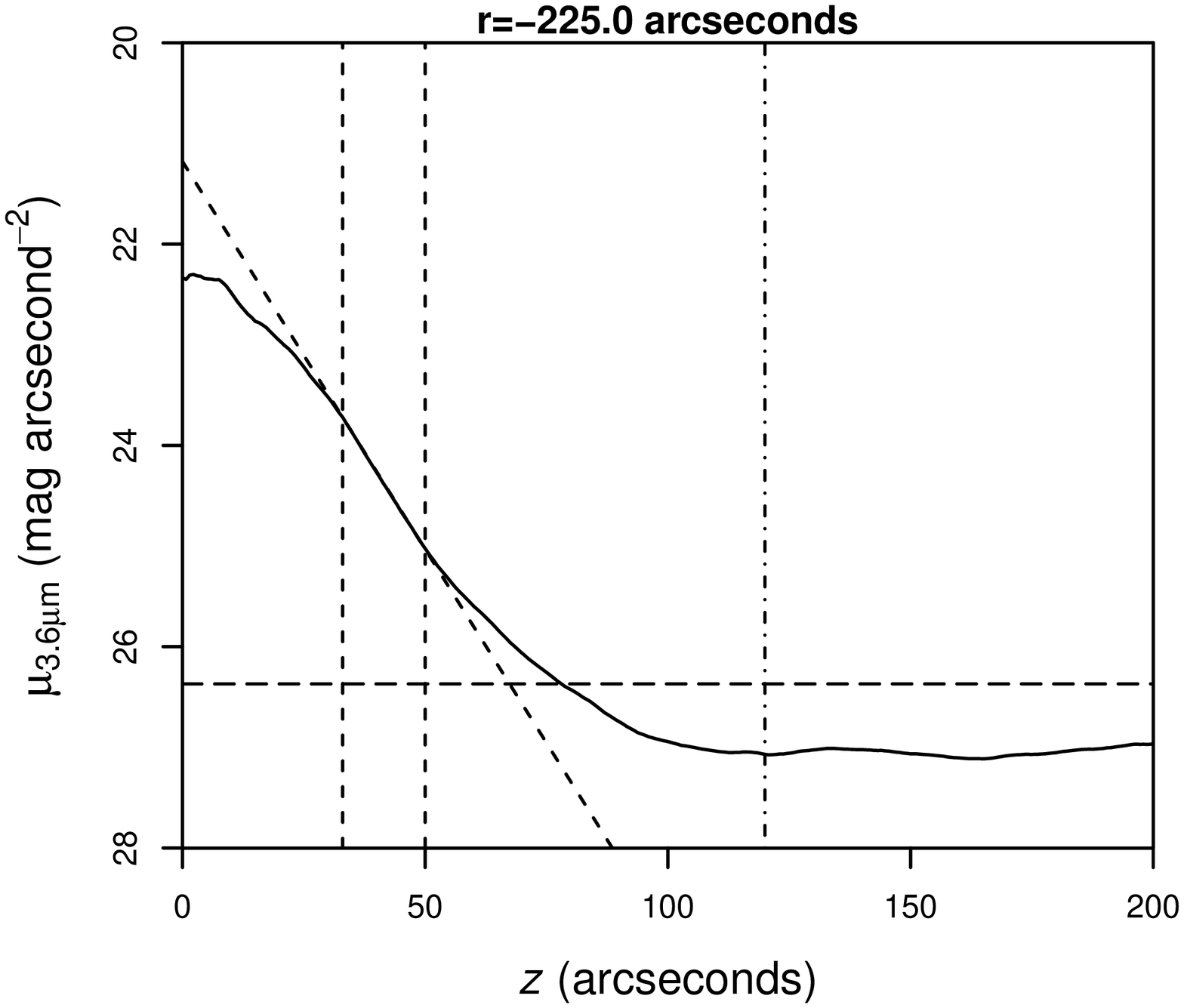}\\

\end{tabular}
\caption{\label{profiles4244} Luminosity profiles along the $z$ axis at different galactocentric radii (seen in the subplot's titles). Short dashed lines indicate the fits to inner exponentials and the range of $z$ at which the fit has been done (vertical lines). Dotted lines indicate the fits to outer exponentials and the range of $z$ at which the fit has been done (vertical lines) for these profiles for which an outer exponential has been fitted (in some cases the vertical line indicating the inner boundary of the outer exponential fitting range is coincident with the outer boundary of the inner exponential fitting range). Dash-dotted lines indicate the minimum $z$ at which the noise level has been calculated. The horizontal long-dashed line indicates the noise level. The profiles from the near side of the galaxy are on the left and those from the far side are on the right hand side. The error bar length at the noise level is on the order of the line width.}
\end{center}
\end{figure*}

\begin{figure*}[!ht]
\begin{center}
\begin{tabular}{c|c}
Near side (bottom side) & Far side (top side) \\
&\\
\includegraphics[width=0.45\textwidth]{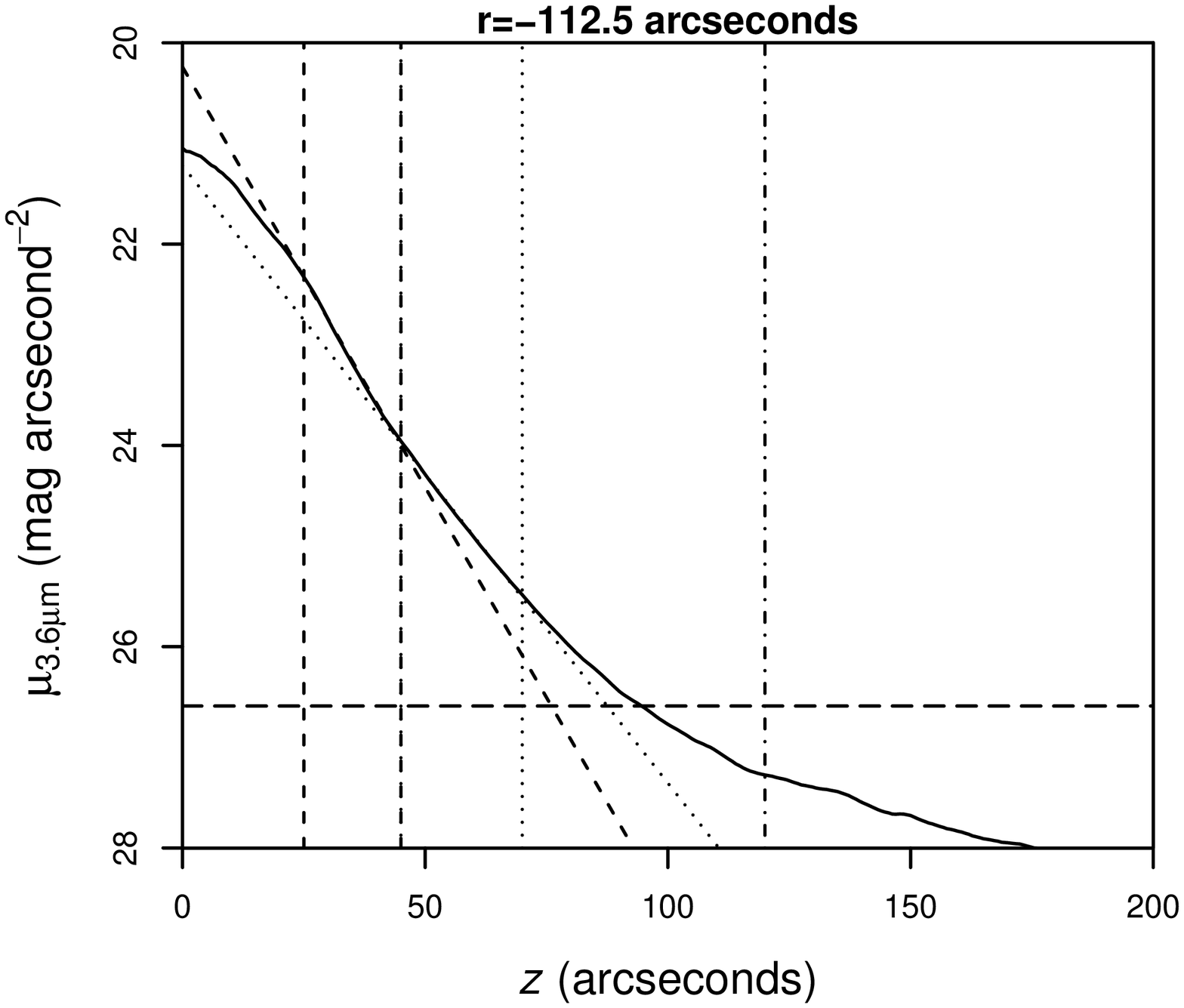}&
\includegraphics[width=0.45\textwidth]{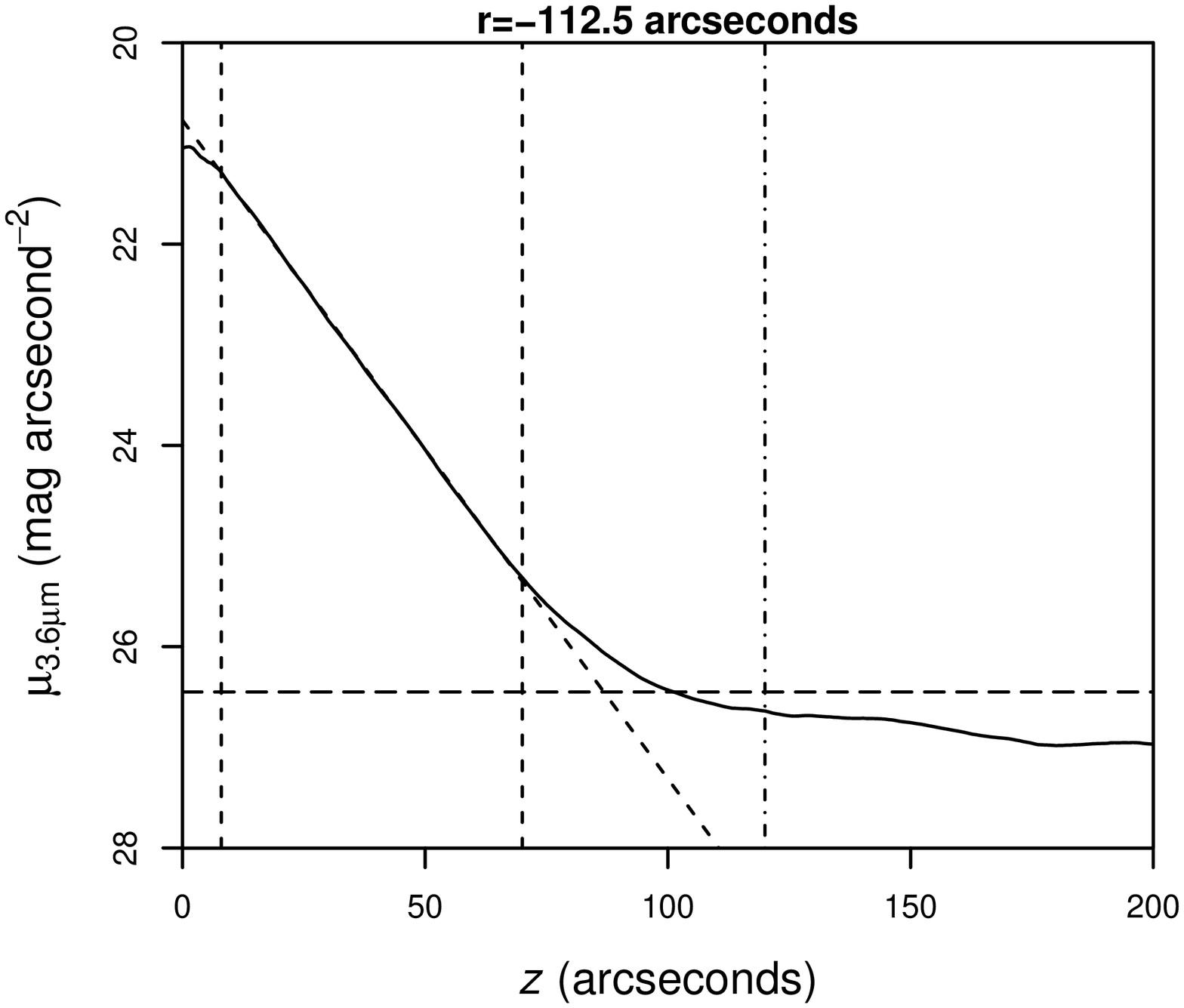}\\
\includegraphics[width=0.45\textwidth]{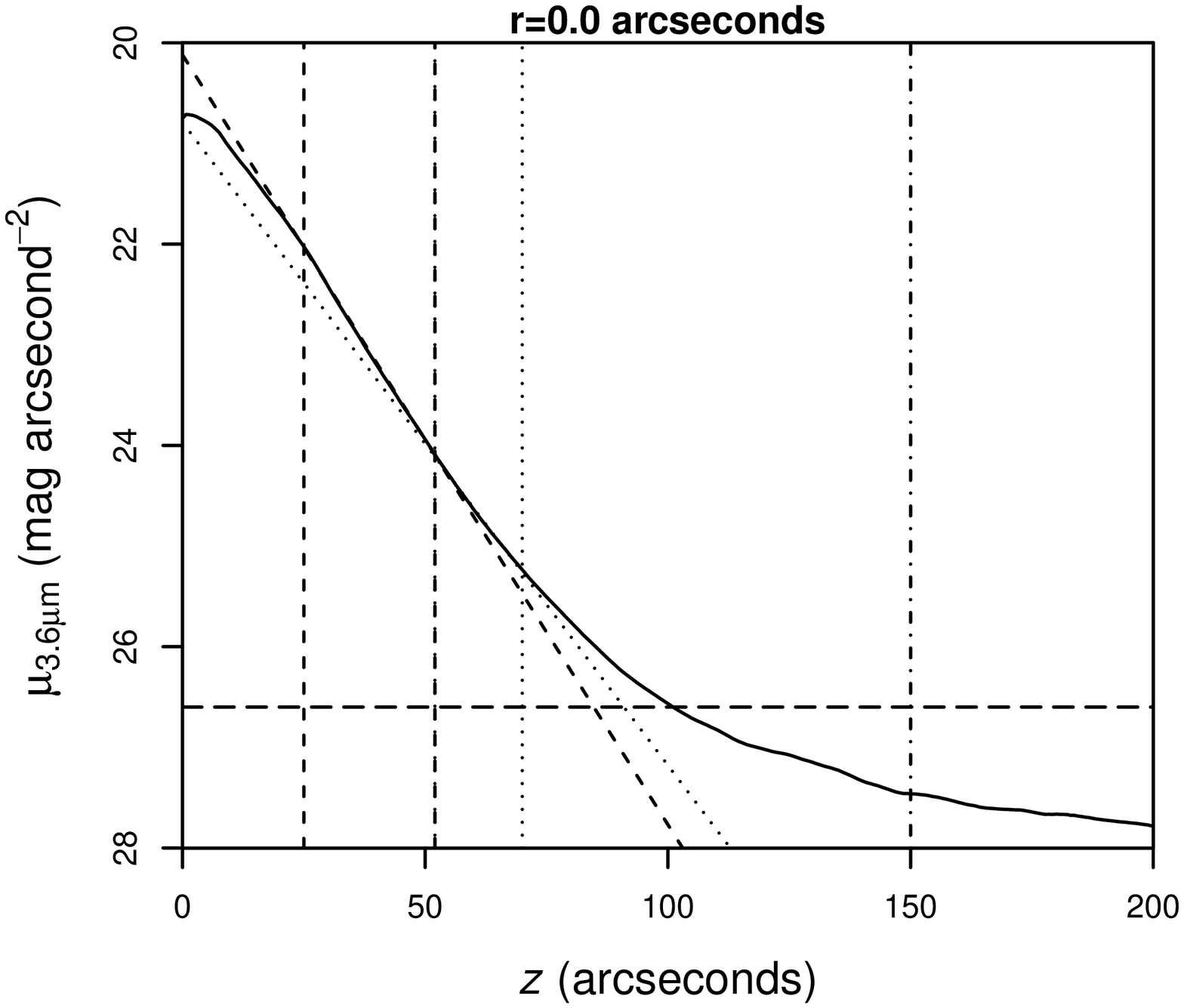}&
\includegraphics[width=0.45\textwidth]{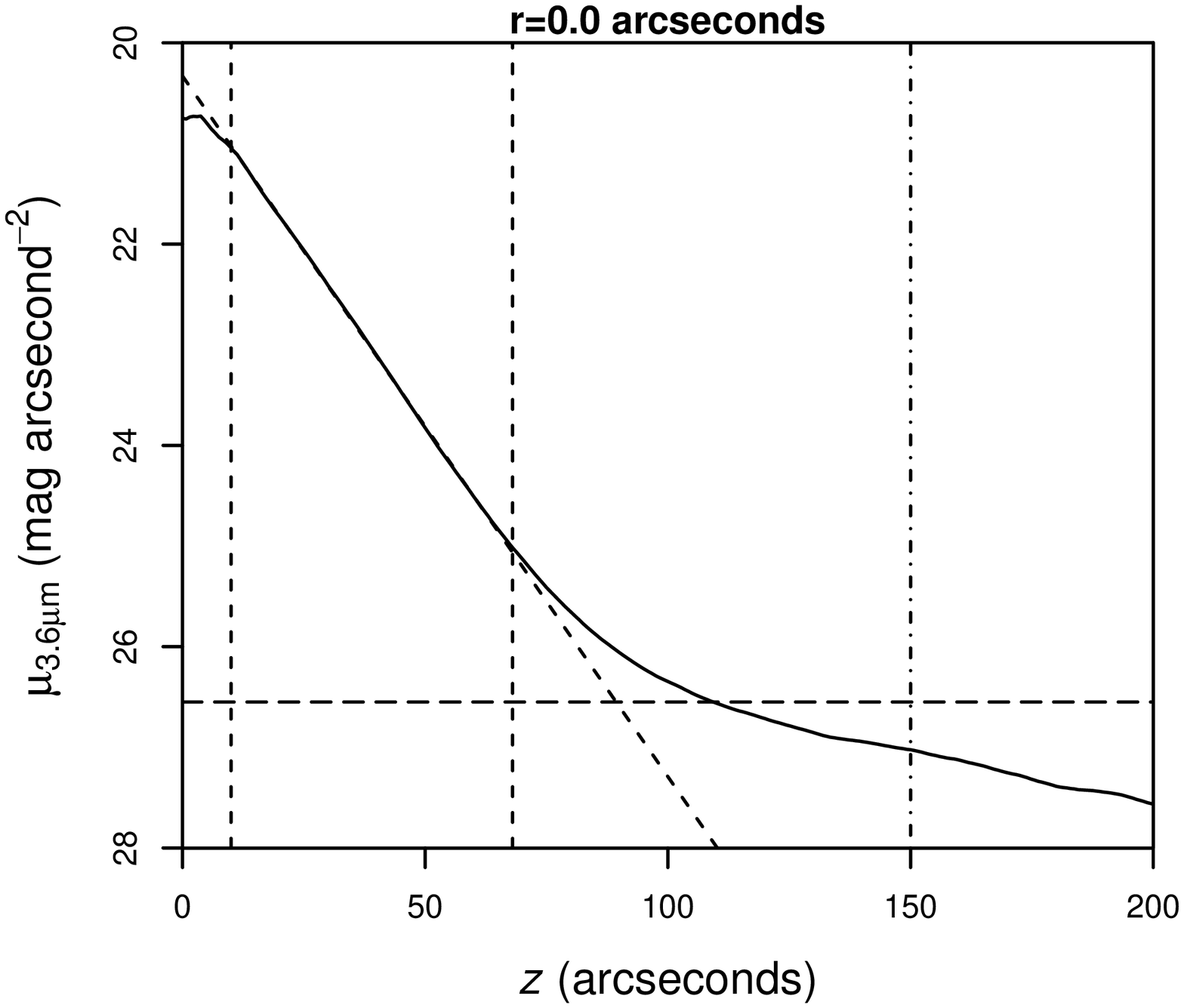}\\
\includegraphics[width=0.45\textwidth]{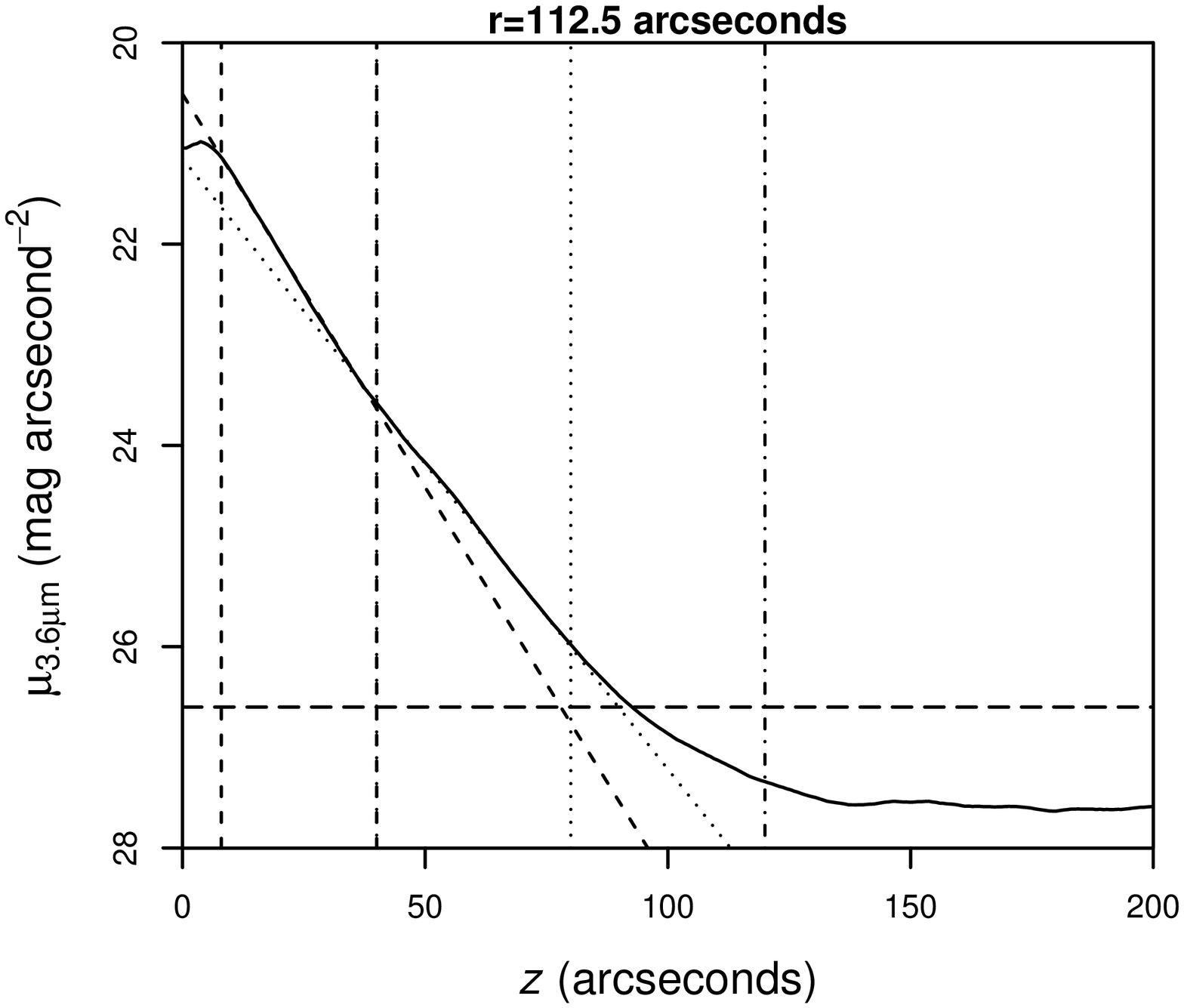}&
\includegraphics[width=0.45\textwidth]{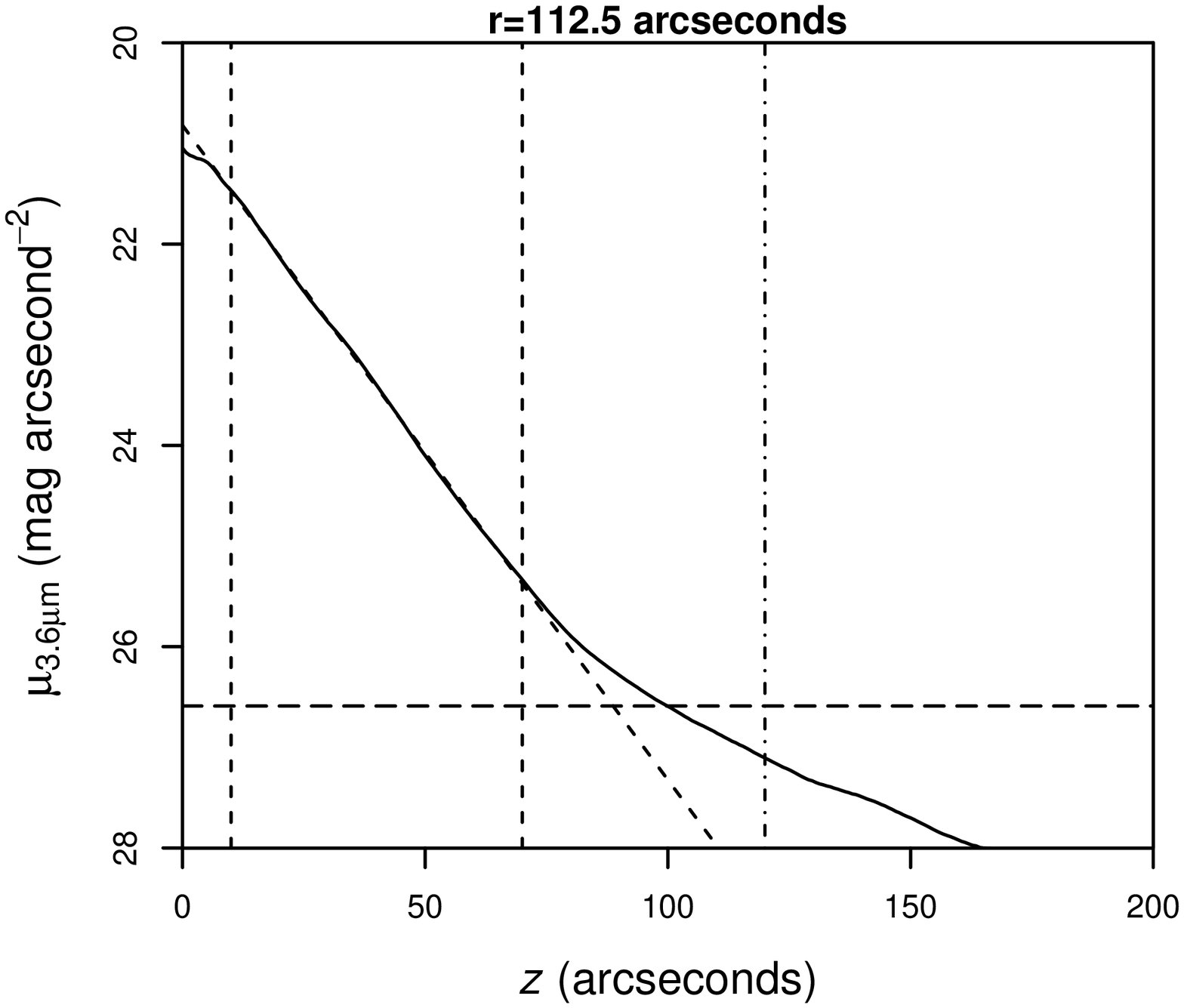}\\
\end{tabular}

{\footnotesize Figure~\ref{profiles4244}: (continued).}
\end{center}
\end{figure*}

\begin{figure*}[!ht]
\begin{center}
\begin{tabular}{c|c}
Near side (bottom side) & Far side (top side) \\

&\\

\includegraphics[width=0.45\textwidth]{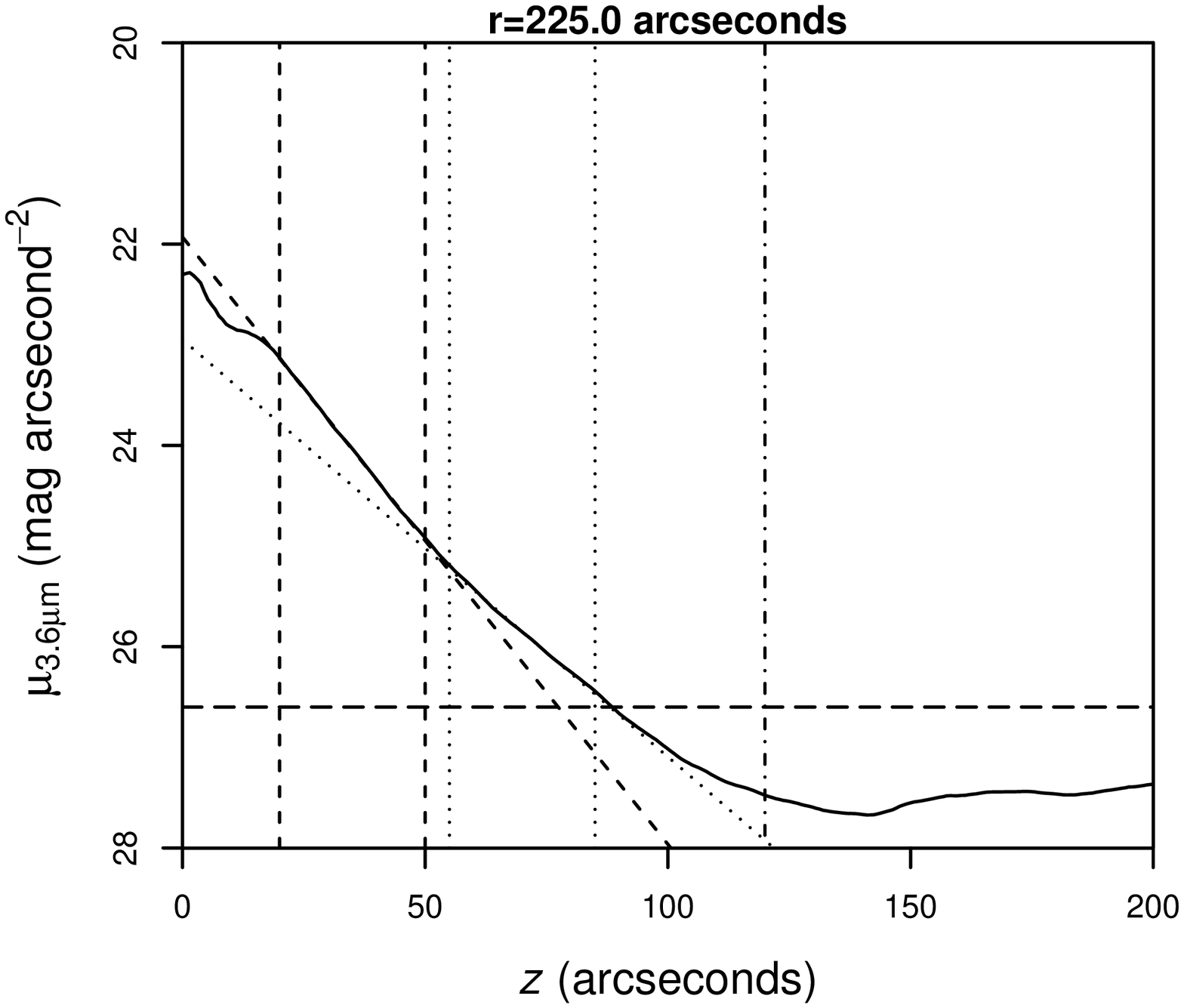}&
\includegraphics[width=0.45\textwidth]{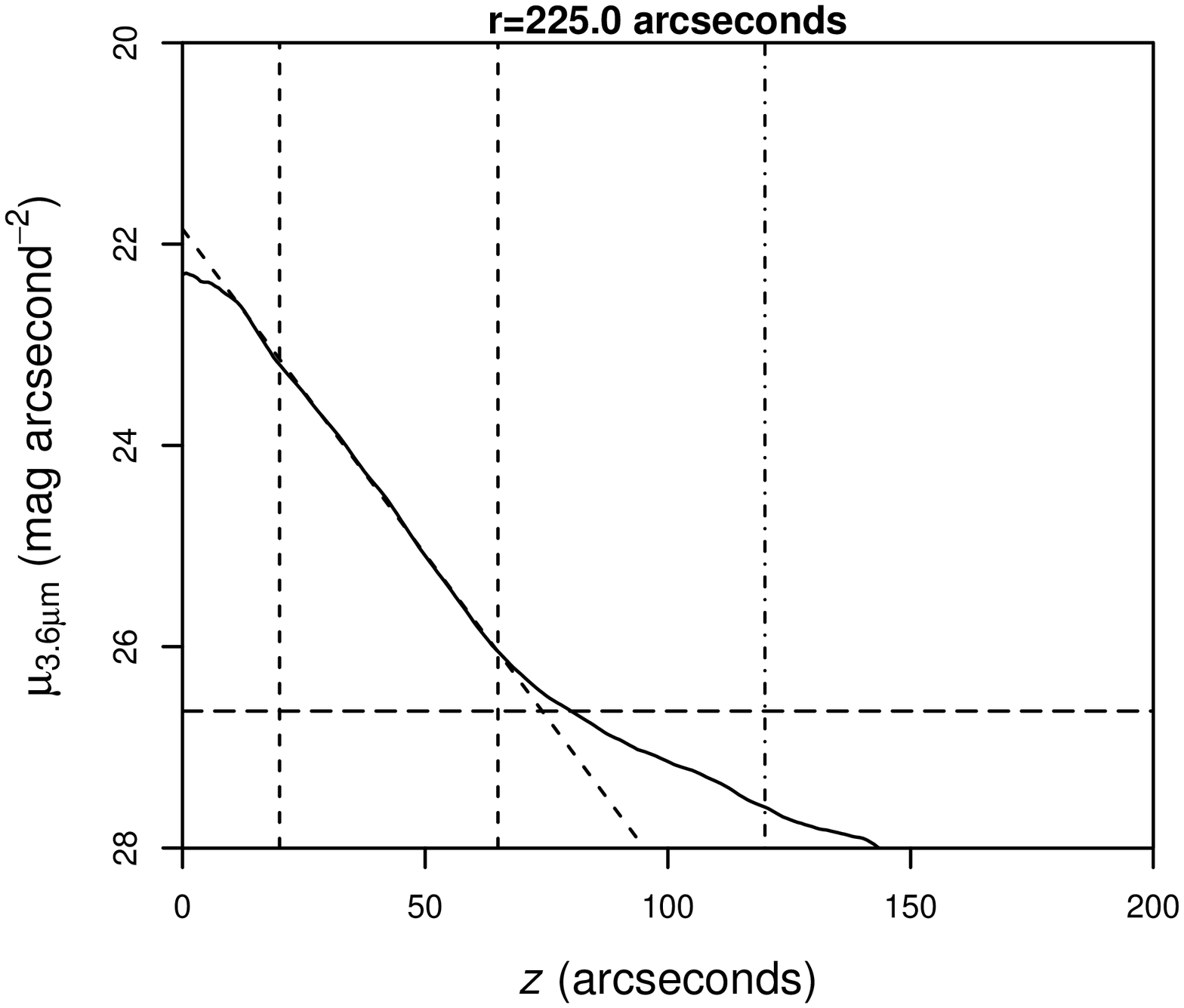}\\
\includegraphics[width=0.45\textwidth]{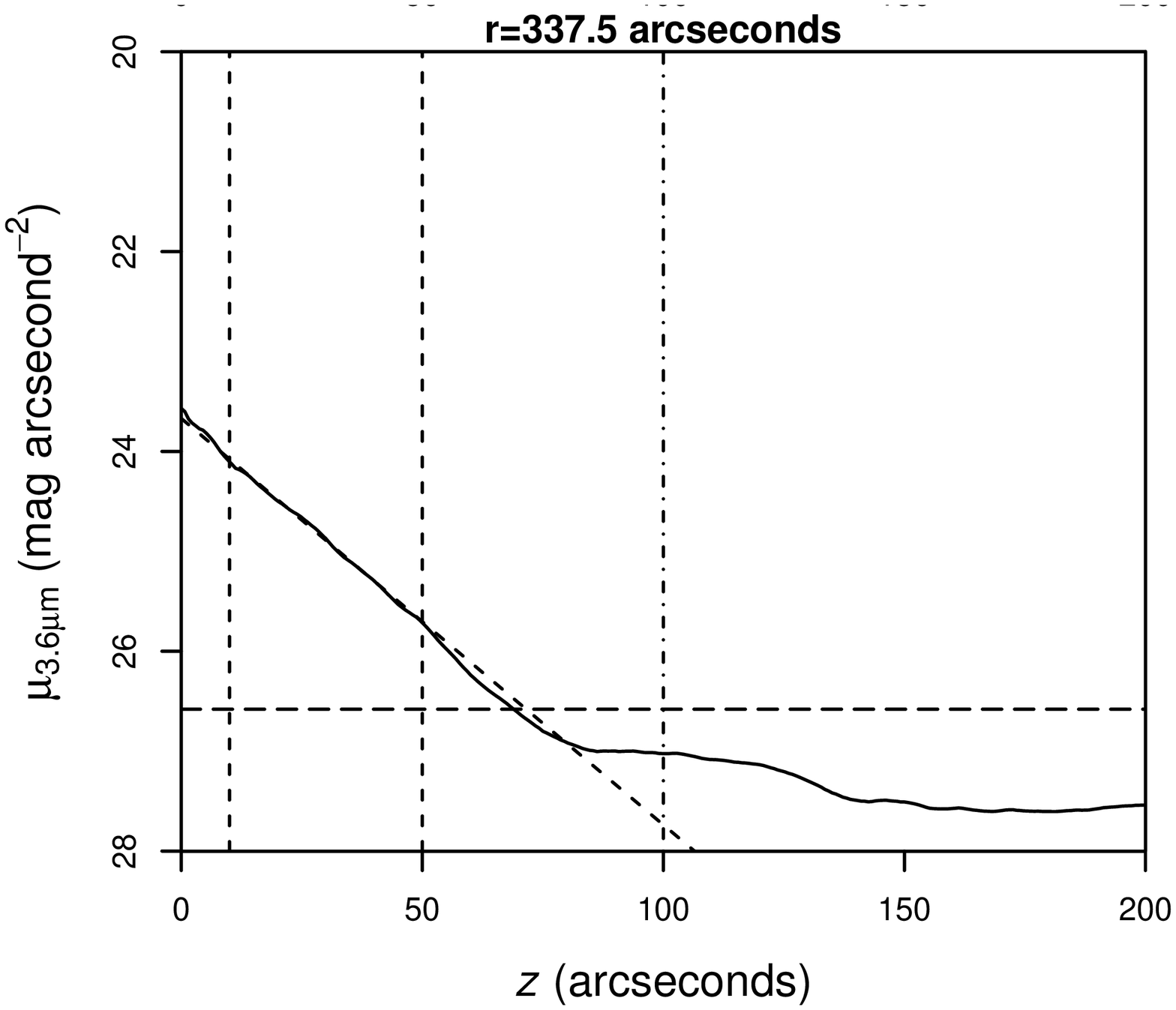}&
\includegraphics[width=0.45\textwidth]{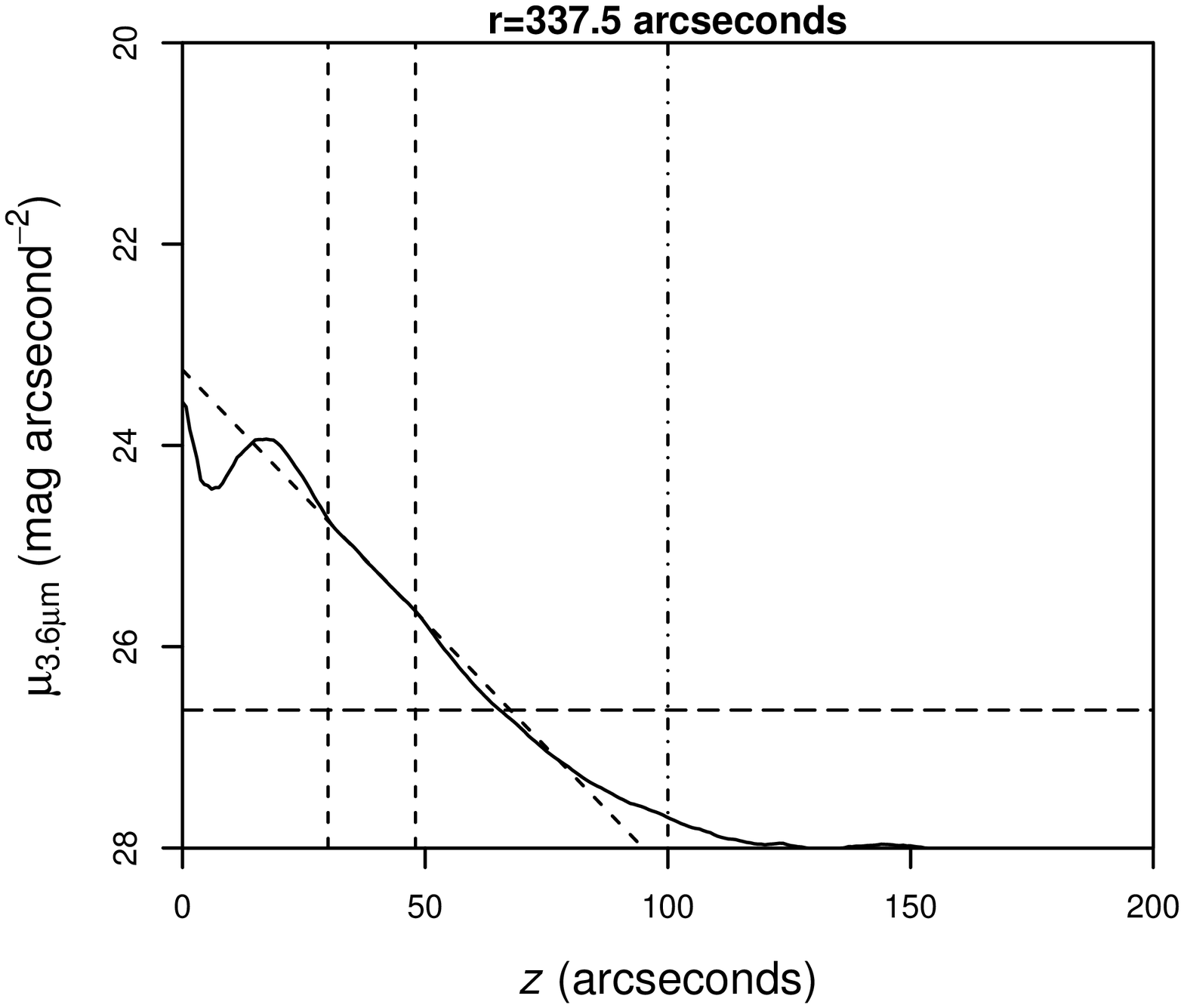}\\
\end{tabular}

{\footnotesize Figure~\ref{profiles4244}: (continued).}
\end{center}
\end{figure*}

\begin{table*}[t]
\begin{center}
\caption[Fitting data of the components in NGC~4244]{\label{fits} Vertical profiles fitting parameters and results.}
\begin{tabular}{c c c | c c c | c c c}
\hline
$r$   & $z_{\rm n}$ & $\sigma$ level &  \multicolumn{3}{c|}{Inner exponential fit} & \multicolumn{3}{c}{Outer exponential fit}   \\
 (\arcsec) &(\arcsec)& (mag\,arcsec$^{-2}$) &range (\arcsec) & $I_{\rm 0t}$ (mag\,arcsec$^{-2}$)& $H_{\rm zt}$ (\arcsec) & range (\arcsec) & $I_{\rm 0T}$ (mag\,arcsec$^{-2}$)& $H_{\rm zT}$ (\arcsec)\\
(1) & (2) & (3) & (4) & (5) & (6) & (7) & (8) & (9) \\
\hline
\multicolumn{8}{c}{Far side profiles}\\
\hline
-337.5 & 100 & 26.32 &$18-60$ & $22.49\pm0.02$ & $18.1\pm0.1$ & $-$ &$-$ & $-$\\
-225.0 & 120 & 26.37 &$33-50$ & $21.18\pm0.02$ & $14.1\pm0.1$ & $-$ &$-$ & $-$\\
-112.5 & 120 & 26.45 &$ 8-70$ & $20.77\pm0.01$ & $16.6\pm0.1$ & $-$ &$-$ & $-$\\
 0.0 & 150 & 26.55 &$10-68$ & $20.33\pm0.01$ & $15.6\pm0.1$ & $-$ &$-$ & $-$\\
 112.5 & 120 & 26.59 &$10-70$ & $20.82\pm0.04$ & $16.7\pm0.1$ & $-$ &$-$ & $-$\\
 225.5 & 120 & 26.64 &$20-65$ & $21.85\pm0.03$ & $16.8\pm0.2$ & $-$ &$-$ & $-$\\
 337.5 & 100 & 26.63 &$30-48$ & $23.25\pm0.03$ & $21.7\pm0.3$ & $-$ &$-$ & $-$\\
\hline
\multicolumn{8}{c}{Near side profiles}\\
\hline
-337.5 & 100 & 26.45 & $18-38$ & $22.18\pm0.03$ & $15.1\pm0.2$ & $38-60$ &$22.47\pm0.05$ & $16.9\pm0.2$\\
-225.0 & 120 & 26.55 & $25-45$ & $21.25\pm0.03$ & $14.2\pm0.1$ & $45-65$ &$21.90\pm0.06$ & $17.5\pm0.3$\\
-112.5 & 120 & 26.59 & $25-45$ & $20.24\pm0.07$ & $13.0\pm0.3$ & $45-70$ &$21.22\pm0.05$ & $17.7\pm0.3$\\
 0.0 & 150 & 26.60 & $25-52$ & $20.12\pm0.03$ & $14.2\pm0.1$ & $52-70$ &$20.79\pm0.09$ & $17.0\pm0.4$\\
 112.5 & 120 & 26.60 & $ 8-40$ & $20.51\pm0.02$ & $13.9\pm0.1$ & $40-80$ &$21.15\pm0.03$ & $17.9\pm0.2$\\
 225.5 & 120 & 26.60 & $20-50$ & $21.93\pm0.02$ & $18.0\pm0.1$ & $55-85$ &$22.95\pm0.06$ & $26.2\pm0.5$\\
 337.5 & 100 & 26.58 & $10-50$ & $23.67\pm0.02$ & $26.7\pm0.4$ & $-$  &$-$   & $-$   \\
\hline
\end{tabular}
\end{center}
Notes: The uncertainties in this table are statistical errors (3$\sigma$). Galactocentric distance (col.~1), $z$ above which noise has been calculated (col.~2), noise level (col.~3), $z$ range at which the fit for the inner exponential has been calculated (col.~4), inner exponential central surface brightness (col.~5), inner exponential scale-length (col.~6), $z$ range at which the fit for the outer exponential has been calculated (col.~7), outer exponential central surface brightness (col.~8), and outer exponential scale-length (col.~9).
\end{table*}

\begin{figure*}[!t]
\begin{center}
\begin{tabular}{c}
\includegraphics[width=0.91\textwidth]{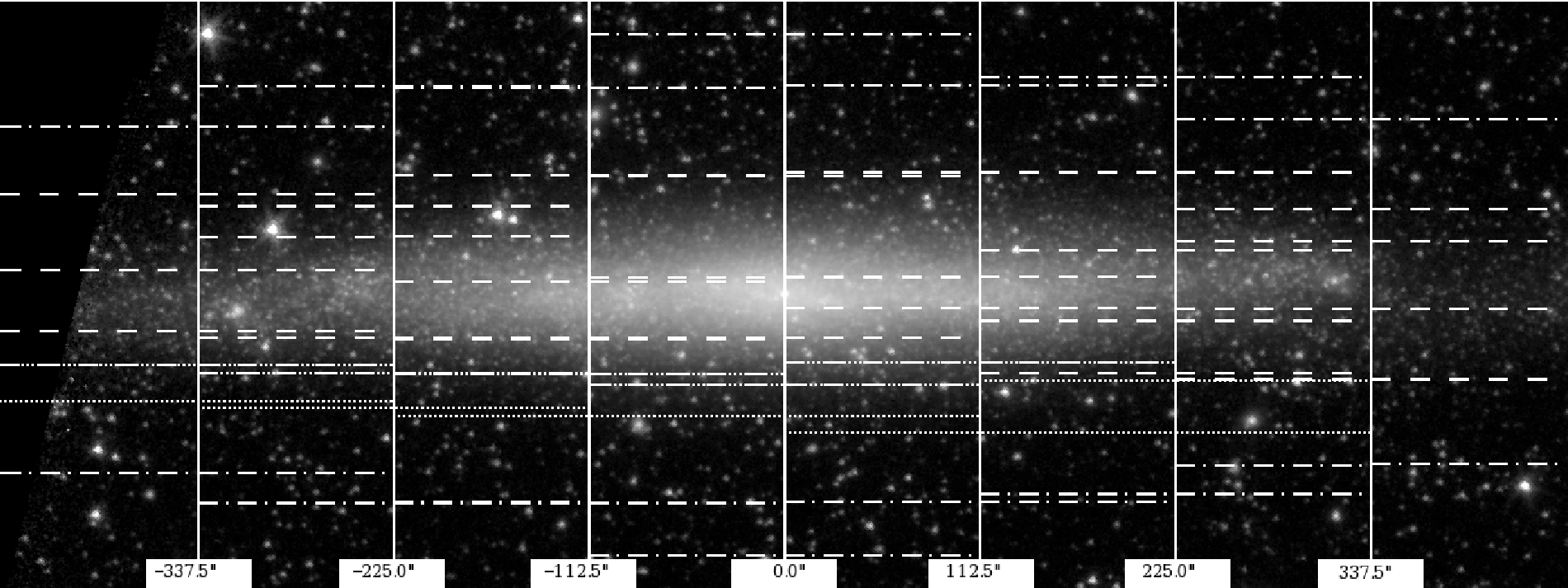}
\end{tabular}
\caption{\label{sections} Representation of the fitting ranges in a $3.6\mu{\rm m}$ image of NGC~4244. The displayed area is the same as in the bottom panel in Fig.~\ref{images4244}. The vertical solid lines denote the center of the bins. As in Fig.~\ref{profiles4244}, short dashed lines indicate the range of $z$ at which the fit to the thin disk has been done. Dotted lines indicate the range of $z$ at which the fit to the thick disk has been done. Dash-dotted lines indicate the minimum $z$ at which the noise level has been calculated. The labels at the center of each bin indicate the bin galactocentric distance, $r$.}
\end{center}
\end{figure*}

\section{Vertical luminosity profiles}

In order to search for traces of a thick disk, we have produced luminosity profiles parallel to the $z$ axis of the galaxy at different galactocentric radii ($r$). These luminosity profiles are shown in Fig.~\ref{profiles4244}.

To produce the luminosity profiles, we first made vertical bins with a width of $225.0\arcsec$ (300\,pixels or 4.8\,kpc), and centered at galactocentric distances $r=-337.5\arcsec$, $-225.0\arcsec$, $-112.5\arcsec$, $0.0\arcsec$, $112.5\arcsec$, $225.0\arcsec$, and $337.5\arcsec$. The $225.0\arcsec$ bin width has been chosen because it offers a good compromise between spatial resolution and the profile's signal-to-noise. As the bin width is double the separation between bins, the bins are overlapping. For each bin we determined the surface brightness at a given $z$ making a median of the counts excluding the pixels which were masked. We determined the location of the mid-plane of the galaxy from the profiles obtained in each bin. This was done by folding the profiles and minimizing the difference between the upper and the lower halves of the light profile, with a precision of 0.1\,pixel. For each bin we made a visual guess of where the galaxy effectively ends. The noise level was calculated by producing the standard deviation of those pixels at higher $z$. This is a conservative estimate of the noise, as the profile is produced by doing the median of the pixels in the bin. We smoothed the profiles in the vertical direction in order to get a higher signal at low surface-brightness areas. This smoothing has been done logarithmically and using a vertical scale-height of 5\,pixels (3.75\arcsec).

As the near and the far side of the galaxy are different, we have fitted them separately. We decided to make exponential fits at ranges in $z$ where the data appear roughly exponential. The fit was made in magnitude units and all the points were given equal weight. We fitted one or two exponentials to each profile depending on whether we detected -- by eye -- any trace of a thick disk or not. The fitting ranges appear in Table~\ref{fits}, and they are indicated in Figs.~\ref{profiles4244} and~\ref{sections}. When preparing the fits, we considered all the luminosity values above the noise level to be reliable. The noise in our profiles is $\mu_{3.6\mu{\rm m}}(AB)\sim26.5\,{\rm mag\,arcsec^{-2}}$. If we had not taken the background gradient into account, as described in Sect.~2, this level would have been $\mu_{3.6\mu{\rm m}}(AB)\sim26.0\,{\rm mag\,arcsec^{-2}}$. The results of the fits are overlaid on the profiles in Fig.~\ref{profiles4244} and are summarized in Table~\ref{fits}.

\section{Results}

 We find that all but one of the far side profiles can be fitted with a single exponential. Only for the profile at $-225.0\arcsec\leq r \leq112.5$ there is an extra faint component which cannot be fitted by an exponential. The near side profiles are well fitted by two exponential components except for that at $r=337.5\arcsec$ (Fig.~\ref{profiles4244}). The scale-heights of all these fits are roughly stable with varying galactocentric radius.

\begin{figure}[!t]
\begin{center}
\begin{tabular}{c}
\includegraphics[width=0.45\textwidth]{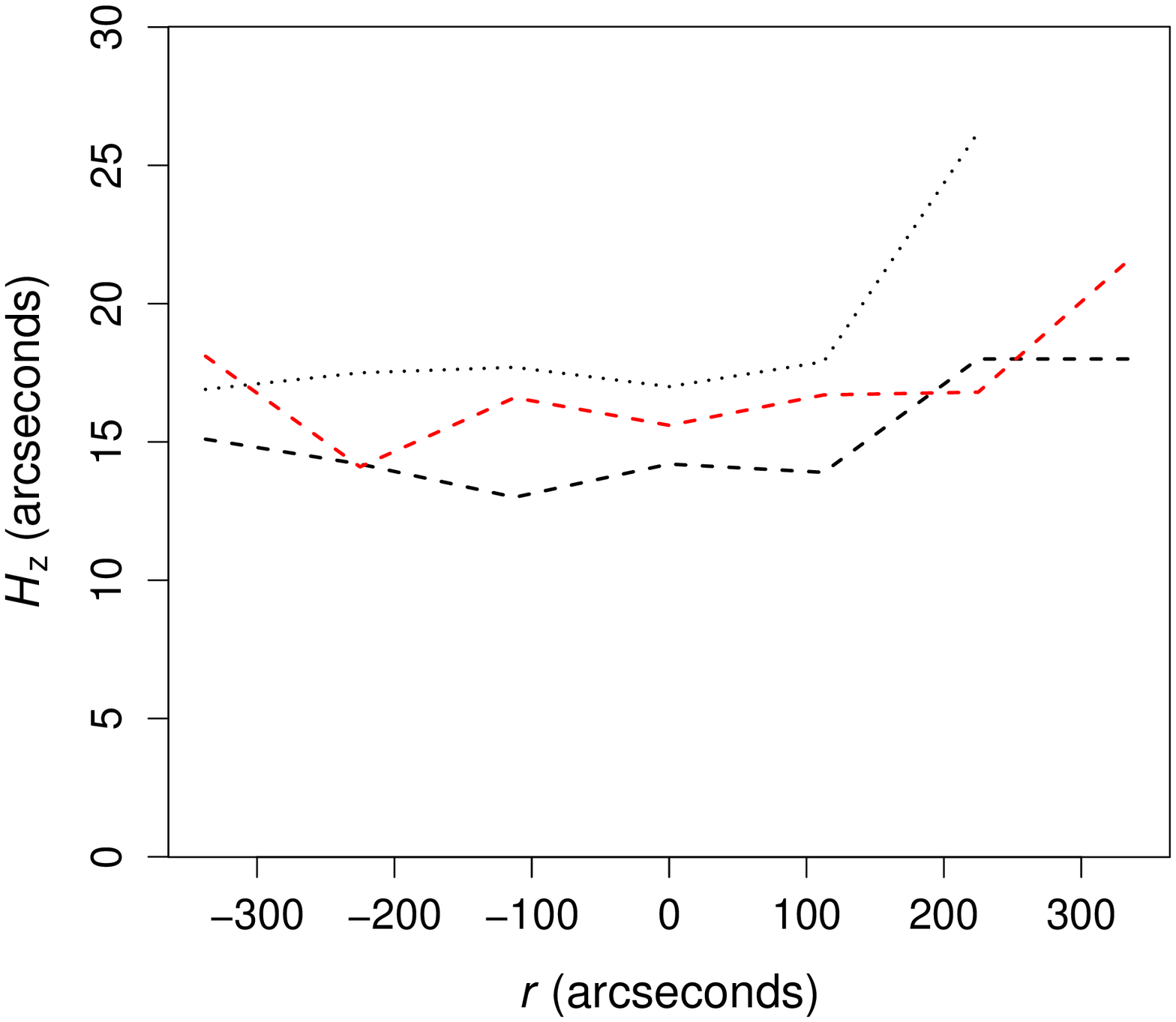}
\end{tabular}
\caption{\label{scaleheight} Scale-height of the disks at different galactocentric radii. Dashed lines indicate the thin disk and the dotted line indicates the thick disk. Black lines indicate the measurements of scale-heights in the near side of the galaxy and red lines the measurements in the far side of it.}
\end{center}
\end{figure}

\subsection{Looking for a thick disk in the luminosity profiles}

The luminosity profiles (Fig.~\ref{profiles4244}) expose a lack of symmetry between the far and the near sides of NGC~4244. Once an inner exponential has been fitted, it is clearly seen that larger $z$ sections of the near side profiles can be fitted with a shallower exponential. By comparing lower and upper profiles, we can see that the concave inflection points between two exponentials in the near side profiles (which are especially noticeable for $-112.5\arcsec$, $112.5\arcsec$, $225.0\arcsec$, and $337.5\arcsec$) have no clear counterpart in the far side part of the galaxy. The inflection point is not so evident in the near side profile at $r=-337.5\arcsec$ and $r=-225.0\arcsec$ and it is not noticeable at all for $r=337.5\arcsec$ may be due to some effect related to this region is being affected by the warp-like feature. The inflection point is also subtle at $r=0.0\arcsec$.

An excess of light appears in the far side of the galaxy, for $-225.0\arcsec\leq r\leq112.5\arcsec$, but it cannot be easily fitted by an exponential and, even where we could in principle have done it, the fit would have to be done over a much smaller range in $z$. This excess corresponds to the arc-like feature in the residual image obtained from {\sc Galfit} (indicated with a white rectangle in the bottom-right panel in Fig.~\ref{modelsngc4244} and shown in Fig.~\ref{arc}). As it has a surface brightness comparable to that of the second exponential we find in the near side profiles, it could hide part of a thick disk in the far side of the galaxy for radii from $r=-225.0\arcsec$ to $r=112.5\arcsec$. However, the lack of traces of a thick disk in the far side profiles, even at galactocentric radii not affected by the non-exponential excess of light, indicates that a thick disk there is less obvious than in the near side of the galaxy, if it is present at all.

The luminosity profiles show some anti-symmetry, which is expected for a more or less symmetric warp or a symmetric spiral arm pattern embedded in the thin disk. This anti-symmetry is seen in Fig.~\ref{profiles4244}, whose panels at large negative radius at the near side resemble most the panels at large positive radius at the far side, and vice-versa.

The scale-height of both disk components is stable from $r=-337.5\arcsec$ to $r=112.5\arcsec$ (Fig.~\ref{scaleheight}). For this range of galactocentric radii, the thin disk scale-height is on average 16.2\arcsec\ (350\,pc) for the far side of the galaxy and 14.1\arcsec\ (300\,pc) for the near side of it. At the same range of galactocentric radii the thick disk has a scale-height of 17.4\arcsec\ (370\,pc). For comparison Seth et al.~(2007) fitted a single disk with $\sim300$\,pc in scale-height. At galactocentric radii larger than $r=112.5\arcsec$, both the thin and thick disks apparently flare (this can also be seen at the right side of Fig.~\ref{dr7} and has been described by van der Kruit \& Searle 1981). This apparent flare is related to the warp-like feature, which we argued may be related to the spiral structure of the disk.

 The definition of scale-height used here is different from that used in other works such as the one by Yoachim \& Dalcanton (2006) and thus the results are not directly comparable. In other papers a two-component model is fitted, but in our case, due to the very subtle change in the slope of the luminosity profile between the thin and thick disk-dominated $z$ ranges, we have considered safer to fit a single exponential in each $z$-range. As a consequence, what we have defined to be the thin disk scale-height is less steep than what would be obtained from a two-component fit as it is influenced by the underlying thick disk component.

\subsection{Thin and thick disks and opacities}

\begin{figure}[!t]
\begin{center}
\begin{tabular}{c}
\includegraphics[width=0.45\textwidth]{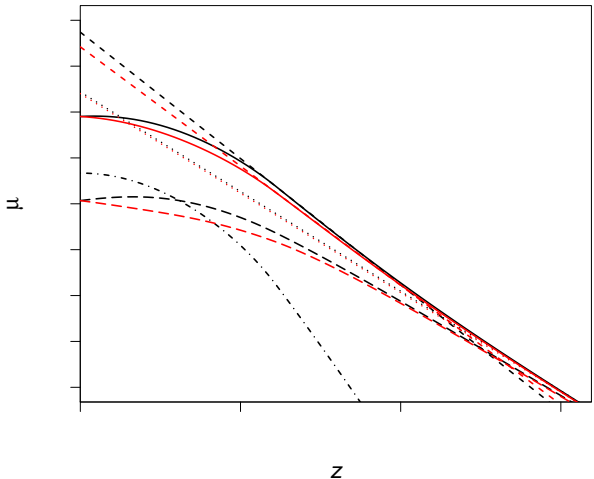}
\end{tabular}
\caption{\label{model} Representation of luminosity profiles of a highly inclined ($i=84.5\deg$) modeled galaxy with a semi-transparent thin disk with an $h_{\rm zT}=2.5h_{\rm zt}$, $I_{\rm 0T}=0.5I_{\rm 0t}$, and a dust distribution in equilibrium with the thin disk component with an opacity set in a way to have an optical depth of $\tau=1$ at $z=0$ at the arbitrary galactocentric radius we are plotting. $\tau$ has been measured through the line which goes from observer eye to the infinity beyond NGC~4244 in the direction we are observing. The far side profile is shown in red and the near side profile is shown in black. Dashed lines indicate the exponential fits to the thin disk and dotted lines indicate the exponential fits to the thick disk. The dash-dotted line indicates the light contribution of the thin disk and the long-dashed lines indicate the light contribution of the thick disk. The tick marks in the vertical axis are drawn every 0.5\,magnitudes. The tick marks in the horizontal axis are separated by a distance equivalent to the fitted thick disk scale-length. $z=0$ has been set to be the mid-plane of the galaxy.}
\end{center}
\end{figure}

We now explore whether the fact that NGC~4244 is not totally edge-on ($i=84.5\deg$ according to Olling 1996), could explain the lack of traces of a thick disk in the far side of the galaxy. In other words, we want to test whether, except for the arc-like feature, the near and far sides of NGC~4244 could be intrinsically symmetric with respect to the mid-plane of the galaxy. To do so we have designed a model galaxy with two disks, each with the following light emission distribution per unit of volume (this light emission distribution can be considered as a mass distribution if we assume a constant mass-to-light ratio):

\begin{equation}
\epsilon(R,z)=\epsilon_{0}\,e^{-R/h_{\rm R}} e^{-|z|/h_{\rm z}}
\end{equation}

\noindent where $\epsilon_{0}$ is the central emission per unit of volume, $R$ the distance to the rotation axis of the galaxy, $z$ the height above its mid-plane, $h_{\rm R}$ the radial light emission distribution scale-length and $h_{\rm z}$ the vertical light emission distribution scale-height. The parameters of each disk will be from now on designated with subindices t for the thin disk and T for the thick disk. $\epsilon_{0}$ should not be confused with $I_{0}$, which is the fitted central surface brightness of the galaxy projected in the sky plane, and $h_{\rm z}$ should not be confused with $H_{\rm z}$, which is the fitted scale-height of the disk in the projected galaxy. Even in the case of a galaxy with no dust and with $i=90\deg$, $H_{\rm z}$ would be different to $h_{\rm z}$, as $h_{\rm z}$ is similar to the definition of scale-height used by Yoachim \& Dalcanton (2006) and $H_{\rm z}$ is the scale-height measured as a single-component exponential fit to a given range in $z$ (see the discussion at the end of Sect.~5.1).

The galaxy model code was designed to have a tunable thick disk central light emission, $\epsilon_{\rm 0T}$, and a tunable thick disk mass distribution scale-height ($h_{\rm zT}$). The disk scale-length was designed to be the same for the thin and the thick disk, which has been found to be true from observations as a first approximation (the ratio between thick and thin disk scale- lengths is on average around 1.25 according to Yoachim \& Dalcanton 2006). We made the scale-height of the thin disk to be one tenth of the scale-length. We made the thin disk semi-transparent, with a local density of dust proportional to the local mass density, which means that the dust is assumed to be in equilibrium with the stellar thin disk component. This assumption is the most important caveat in our modeling, as we do not know which is the real distribution of diffuse dust in NGC~4244. This implies that the results in this section are only qualitative. We assume that the thick disk is transparent, i.e., contains little or no dust. The density of dust in the thin disk is a tunable parameter in our model. Our code projects the light emitted by our three-dimensional model galaxy using a tunable inclination angle, $i$. We have studied our model at $r$ equal to one scale-lenght of the disks.

In order to get vertical luminosity profiles similar to that in Fig.~\ref{profiles4244} at the galactocentric radii at which a thick disk is detected, we have manually tuned the variable parameters of our model, except the galaxy inclination, which we have set to be $i=84.5\deg$ (Olling 1996). We find that a good agreement is obtained when $\epsilon_{\rm 0T}=0.5\epsilon_{\rm 0t}$, $h_{\rm zT}=2.5h_{\rm zt}$, and the dust density is set to have an optical depth $\tau=1$ when $z=0$ at the arbitrary galactocentric radius in which we are making the luminosity profile (see Fig.~\ref{model}). The selected $\tau$ is on the order of that deduced for the mid-plane of an edge-on galaxy in the $K$-band by Bianchi (2008) and thus is a reasonable upper limit for the mid-plane obscuration at $3.6\mu{\rm m}$. All $\tau$ values in this paper have been measured through the line which goes from the observer to infinity, in the direction of NGC~4244. Using these parameters, we find that for the bottom parts of the galaxy the ratio between the thick and the thin disk scale-height of the integrated distribution of light along the line of sight is $H_{\rm zT}\sim1.25H_{\rm zt}$, and that the fitted central surface brightness of the thick disk is between 0.5 and 1.0 magnitudes dimmer than that of the thin disk, which is compatible with values obtained from the fits in Fig.~\ref{profiles4244}.

In Fig.~\ref{profiles4244}, as in our model, the luminosity profiles generally show a steeper thin disk scale-height for the near side of the galaxy ($300\pm15$\,pc against $350\pm30$\,pc when the scale-height is averaged between $r=-337.5\arcsec$ to $r=112.5\arcsec$; the uncertainty corresponds to one standard deviation of the values used for the average scale-height computation). This is because for the far side of the galaxy, the thin disk opacity is obscuring the thick disk, thus removing light from the regions at low $z$, and making the slope in these regions (assigned to the thin disk in our fit) similar to that at higher $z$. This effect is seen in lines representing the thick disk luminosity profile in Fig.~\ref{model}.

In addition, for galactocentric radii between $r=-225.0\arcsec$ and $r=112.5\arcsec$, the arc-like feature may be hiding any trace of a softened transition from the thin to the thick disk in the far side of the galaxy. At galactocentric radii larger than $r=225.0\arcsec$, the galaxy disk starts to be distorted by, possibly, the presence of spiral arms, thus making the comparison between NGC~4244 and the model more difficult.

It can be argued that $\tau=1$ is a very high opacity for a galaxy imaged in $3.6\mu{\rm m}$, but this opacity only corresponds to the mid-plane of the galaxy, where the central dust lane is found. $\tau$ drops fast with $z$, and it is on the order of $\tau=0.1$ at the $z$ at which the transition between the thin and the thick disk is found. In addition, we find that there is a degeneracy between the opacity and $\epsilon_{0T}/\epsilon_{0t}$. As an example, setting the mid-plane opacity as $\tau=0.6$, and $\epsilon_{0T}=0.65\epsilon_{0t}$ yields a luminosity profile very similar to that in Fig.~\ref{model}.

\begin{figure}[!t]
\begin{center}
\begin{tabular}{c c}
\includegraphics[width=0.45\textwidth]{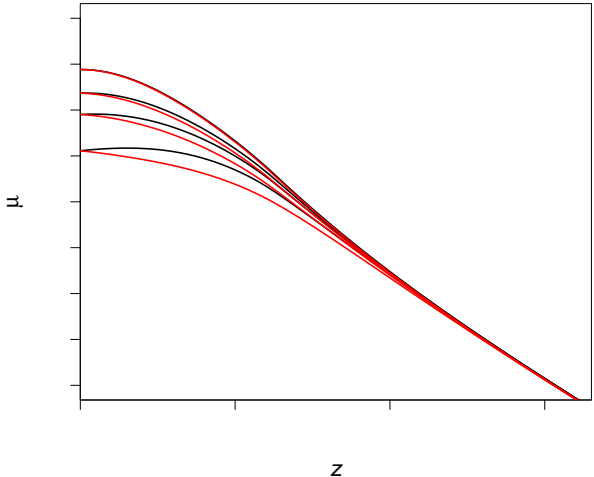}
\end{tabular}
\caption{\label{model_tau} Representation of luminosity profiles of a highly inclined ($i=84.5\deg$) modeled galaxy with a semi-transparent thin disk with $h_{\rm zT}=2.5h_{\rm zt}$, $I_{\rm 0T}=0.5I_{\rm 0t}$, and a dust distribution in equilibrium with the thin disk component with an opacity set in a way to have an optical depth of $\tau=0.00$, $\tau=0.50$, $\tau=1.00$, and $\tau=2.00$ (lines from up to down) at $z=0$ and at the arbitrary galactocentric radius we are plotting. $\tau$ has been measured through the line which goes from observer eye to the infinity beyond NGC~4244 in the direction we are observing. The far side profile is shown in red and the near side profile is shown in black. The tick marks in the vertical axis are drawn every 0.5\,magnitudes. The tick marks in the horizontal axis are separated by a distance equivalent to the thick disk scale-length obtained from the fit in Fig.~\ref{model}. $z=0$ has been set to be the mid-plane of the galaxy.}
\end{center}
\end{figure}

We have tested, in the case of a galaxy with $i=84.5\deg$, $\epsilon_{0T}=0.5\epsilon_{0t}$ and $h_{\rm zT}=2.5h_{\rm zt}$ the behavior of the luminosity profiles as a function of the amount of dust in the disk. The result is shown in Fig.~\ref{model_tau} in which we see that for a mid-plane optical depth of $\tau=0$, both sides are symmetric. When increasing $\tau$, the central luminosity is reduced for both near and far side profiles. The reduction in the central surface brightness causes $H_{zt}$ (the measured scale-height of the thin disk) to be increased, making the change in slope between the thin and thin disk softer with increasing $\tau$. This effect is more pronounced for the far side profile. The trend described in this paragraph is general and also applies when studying highly-inclined disks with different $\epsilon_{0T}/\epsilon_{0t}$ and $h_{\rm zT}/h_{\rm zt}$.

\begin{figure}[!t]
\begin{center}
\begin{tabular}{c c}
\includegraphics[width=0.45\textwidth]{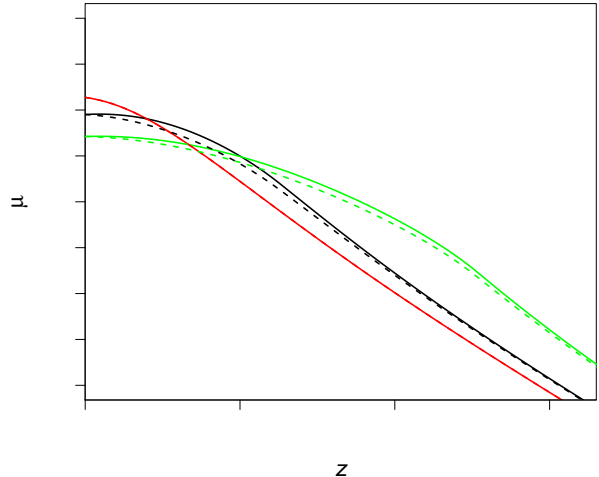}
\end{tabular}
\caption{\label{model_angle} Representation of luminosity profiles of modeled galaxies with different inclinations ($i=90.0\deg$, $i=84.5\deg$, and $i=79.0\deg$ which are represented in red, black, and green respectively) with a semi-transparent thin disk with an $h_{\rm zT}=2.5h_{\rm zt}$, $I_{\rm 0T}=0.5I_{\rm 0t}$, and a dust distribution in equilibrium with the thin disk component with an opacity set in a way to have an optical depth of $\tau=1.00$ when $i=84.5\deg$ at $z=0$ and at the arbitrary galactocentric radius we are plotting. $\tau$ has been measured through the line which goes from observer eye to the infinity beyond NGC~4244 in the direction we are observing. The near side profiles are shown at left and the far side profiles are shown at right. Solid lines indicate the near side and dashed lines indicate the far side. The tick marks in the vertical axis are drawn every 0.5\,magnitudes. The tick marks in the horizontal axis are separated by a distance equivalent to the thick disk scale-length obtained from the fit in Fig.~\ref{model}. $z=0$ has been set to be the mid-plane of the galaxy.}
\end{center}
\end{figure}

We have also tested the effect of varying the galaxy inclination, $i$. The results are displayed in Fig.~\ref{model_angle}. We see that for our modeled galaxy, in the cases of far side luminosity profiles the effect of a thick disk is not enhanced due to inclination effects (the ratio of scale-heights between the thin and thick disk is constant with inclination; for the model with a $\tau=1$ at $z=0$ and $i=84.5\deg$, $H_{\rm zT}/H_{\rm zt}=1.20$ for $i=90.0\deg$ and $H_{\rm zT}/H_{\rm zt}=1.20$ for $i=84.5\deg$), but that in the case of near side luminosity profiles small angle differences with respect to the edge-on position make the difference between thin and thick disk scale-heights larger (for the model with a $\tau=0$ at $z=0$ and $i=84.5\deg$, $H_{\rm zT}/H_{\rm zt}=1.20$ for $i=90.0\deg$ and $H_{\rm zT}/H_{\rm zt}=1.27$ for $i=84.5\deg$). Larger angle differences with respect to the edge-on position makes any trace of thick disk to disappear.

Our results in this Section confirm that the luminosity profiles of NGC~4244 are compatible with the galaxy having a thin and a thick disk and which are symmetric with respect to its mid-plane except for an arc-like feature on its far side.

\section{Summary of NGC~4244 features}

The features we have identified in NGC~4244 are:

\begin{itemize}
\item An unresolved nuclear cluster studied by Seth et al.~(2008).
\item Two disk components, whose identification can only be done in the near side of the galaxy, possibly due to the presence of diffuse dust in the thin disk.
\item Possibly a spiral arm structure which causes the disk to have a slightly `warped' appearance at around 5\arcmin\ from the nucleus.
\item Signs of a possible tidal feature.

\end{itemize}

NGC~4244 also has an \hi\ disk which is coplanar with the optical and near-IR one except at very large galactocentric radii ($r>10$\arcmin; seen in Olling 1996) where it shows a warp going in the opposite direction to the warp-like feature seen at $\sim4\arcmin-5\arcmin$ in galactocentric distance.

\section{Discussion and Conclusions}

We find evidence of two disk components in a galaxy which was reported to have only one or as being a doubtful case. Using modeling we find that the fact of the thick disk is not detected in the far side of the galaxy can be explained by the effect of dust in a not perfectly edge-on galaxy.

Fry et al.~(1999) did not detect the thick disk in their surface photometry because although the $R$-band image has a limiting surface brightness similar to ours, they assumed that the galaxy was symmetric with respect to the mid-plane and they summed the top and bottom luminosity profiles, which diluted the effect of the subtle thick disk. In addition $R$-band is more sensitive to dust, making such a subtle thick disk hard to detect in this band. Seth et al.~(2005b) deduced the presence of an extended component other than the thin disk from the scale-height of the oldest stars (RGB), which have a slight overdensity above the fitted sech$^{-2}$ profile. However, this component is more extended than the thick disk we found and may be a halo. The sum of all stellar components they have studied (young main sequence stars, Helium-burning stars, AGB stars, and RGB stars) shows some trace of what could be a thick disk in the online color version of their Fig.~3. Tikhonov \& Galazutdinova (2005) found an extended feature by counting RGB stars at the location of the arc-like feature. Seth et al.~(2007) studied the RGB star distribution in a thin strip along the bottom part of the minor axis of NGC~4244 deep enough to find a halo component undetected by us due to its extreme faintness, but they did not notice any trace of the thick disk. This is probably due to the fact that the evidence for a thick disk is more subtle at $r=0\arcsec$ than at other galactocentric radii (see Fig.~\ref{profiles4244}). The scale-height for the disk they fit ($\sim300$\,pc or $\sim14\arcsec$) is exactly the same as the one we find for the thin disk at $r=0\arcsec$.

One possible thick disk formation mechanism is dynamical heating of a thin disk. According to recent modeling (Bournaud et al.~2007; 2009), the thick disk should, at least in part, form due to gravitational instabilities. These are also responsible for the creation of giant star-forming clumps which will eventually merge to form the galaxy bulge during the first gigayear after the galaxy formation. As NGC~4244 has virtually no bulge and a thick disk which is not very differentiated from the thin disk, it is conceivable that for some reason the disk dynamical instabilities have been suppressed, which would lead to a lack of thin/thick disk differentiation. The other kind of dynamical heating is a consequence of the encounters of stars with disk substructure and is considered to be a secular evolutionary process due to thin disk irregularities such as spiral arms and giant molecular clouds, or due to dark matter subhaloes. The lack of a very prominent thick disk would indicate that NGC~4244 has got a very smooth structure during most of its life. This smooth structure would have led to a modest `blurring' in the disk of the galaxy. Alternatively, NGC~4244 could have been formed much more recently than other disk galaxies studied so far.

Part of the thick disk could also have formed through accretion of stellar material from merging satellites at the time of the galaxy build-up (Yoachim \& Dalcanton 2006). According to this model the accreted gas builds the thin disk. In this formation scenario, the lack of an extended thick disk would imply that the merging proto-galaxies did not have a substantial stellar component at the time of the formation of NGC 4244.

 Last, but not least, the thick disk may have formed {\it in situ}. According to Elmegreen \& Elmegreen (2006) edge-on galaxies with redshifts $z=0.5-4.5$ (corresponding to look-back times of $4.8-12.2$\,Gyr) have disk morphologies whose ratio between long and short axes are compatible with those of local thick disks. However, these disks should shrink when the galaxy accretes gaseous material, which makes the potential well deeper. In the context of this thick disk origin theory NGC~4244 may be a case of a galaxy having accreted more gaseous material than average, thus making the thick disk thinner than for most disk galaxies. This material accretion must have been through gentle gas inflows which have settled in the thin disk.

To sum up, we have found evidence for two disk components in NGC~4244. The reason for the low degree of differentiation between the thin and thick disk remains unknown and may be related to a lack of strong secular evolution, to a lack of stellar material in the fragments which built up the galaxy, or to a high inflow rate of gaseous material after the galaxy formation.

\section*{Acknowledgments}

The authors wish to thank the entire S$^4$G team for their efforts in this project. We thank our anonymous referee for giving useful advice which have improved the quality of this Paper. We thank Inma Mart\'inez-Valpuesta for her useful comments. We are grateful to the dedicated staff at the Spitzer Science Center for their help and support in planning and execution of this Exploration Science program. We also gratefully acknowledge support from NASA JPL/Spitzer grant RSA 1374189 provided for the S$^4$G project. KMD is supported by an NSF Astronomy and Astrophysics Postdoctoral Fellowship under award AST-0802399. EA and AB thank the Centre National d'\'Etudes Spatiales for financial support.

\end{document}